\documentclass[twocolumn,preprintnumbers,amssymb,aps,prx]{revtex4-1}
\usepackage{graphicx}
\usepackage{dcolumn}
\usepackage{bm}
\usepackage{amssymb}
\usepackage{amsmath}
\usepackage{supertabular}
\usepackage{longtable}
\usepackage{multirow}
\usepackage{tipa}
\usepackage{appendix}
\usepackage{ulem}
\usepackage{xcolor}
\usepackage{tabularx,booktabs}
\usepackage[colorinlistoftodos]{todonotes}
\presetkeys{todonotes}{inline,backgroundcolor=yellow}{}

\begin{document}

\title{Assessing the Precision of Quantum Simulation of Many-Body Effects in Atomic Systems using the Variational Quantum Eigensolver Algorithm}
\author{$^{1,2}$Sumeet, $^2$V. S. Prasannaa, $^{2,3}$B. P. Das and $^4$B. K. Sahoo}
\affiliation{$^1$Qu $\mathit{\&}$ Co B.V., Palestrinastraat 12H, 1071 LE Amsterdam, The Netherlands \\ $^2$Centre for Quantum Engineering, Research and Education, TCG CREST, Salt Lake, Kolkata 700091, India \\ 
$^3$Department of Physics, Tokyo Institute of Technology, 2-12-1-H86 Ookayama, Meguro-ku, Tokyo 152-8550, Japan\\
$^4$Atomic, Molecular and Optical Physics Division, Physical Research Laboratory, Navrangpura, Ahmedabad 380009, India}

\date{\today}

\begin {abstract}
The emerging field of quantum simulation of many-body systems is widely recognized as a very important application of quantum computing. A crucial step towards its realization in the context of many-electron systems requires a rigorous quantum mechanical treatment of the different interactions. In this pilot study, we investigate the physical effects beyond the mean-field approximation, known as electron correlation, in the ground state energies of atomic systems using the classical-quantum hybrid variational quantum eigensolver (VQE) algorithm. To this end, we consider three isoelectronic species, namely Be, Li$^-$, and B$^+$. This unique choice spans three classes, a neutral atom, an anion, and a cation. We have employed the unitary coupled-cluster (UCC) ans\"{a}tz to perform a rigorous analysis of two very important factors that could affect the precision of the simulations of electron correlation effects within a basis, namely mapping and backend simulator. We carry out our all-electron calculations with four such basis sets. The results obtained are compared with those calculated by using the full configuration interaction, traditional coupled-cluster and the UCC methods, on a classical computer, to assess the precision of our results. A salient feature of the study involves a detailed analysis to find the number of shots (the number of times a VQE algorithm is repeated to build statistics) required for calculations with IBM Qiskit's QASM simulator backend, which mimics an ideal quantum computer. When more qubits become available, our study will serve as among the first steps taken towards computing other properties of interest to various applications such as new physics beyond the Standard Model of elementary particles and atomic clocks using the VQE algorithm. 
\end{abstract}

\maketitle

\section{Introduction}\label{sec1}

Recent advances in quantum information science and technology have heralded the second quantum revolution~\cite{sqr}. These developments have led to new pathways to tackle the challenging quantum many-body problem using quantum computers and simulators~\cite{rev, fhm, water, babbush, ham, dlevel, mogvqe,sqr}. The interest in many-body aspects of electronic structure using quantum computers/simulators stems from the potential speed-up that a quantum computer promises to offer~\cite{AandL1,AandL2,TS} over a classical computer (ClC) in calculating properties such as energies. An overview of the developments in this field can be found Ref.~\cite{rev}. Among the algorithms that calculate the ground state energy of a quantum many-body system, approaches such as the quantum phase estimation algorithm~\cite{AandL2,QPE} may produce energy estimates with high accuracy, but require long coherence times~\cite{lanyon, Mh, bay}. An alternative that promises to alleviate this problem, especially in the noisy-intermediate scale quantum (NISQ) era that we are now in, is the Variational Quantum Eigensolver (VQE) algorithm~\cite{mhyung,peruzzo}. The underlying principle of VQE is to minimize the ground state energy of a system through a quantum-classical hybrid approach by tuning the variational parameters in the appropriate quantum circuit. It has been experimentally realized in platforms such as photonic processors~\cite{peruzzo}, superconducting qubits~\cite{superq}, ion traps~\cite{trappedionq}, etc.. 

\begin{figure*}[t]
\centering
\begin{tabular}{@{\hskip -0.0in}c@{\hskip 0in}c}
    \includegraphics[width=9.0cm, height=9.2cm]{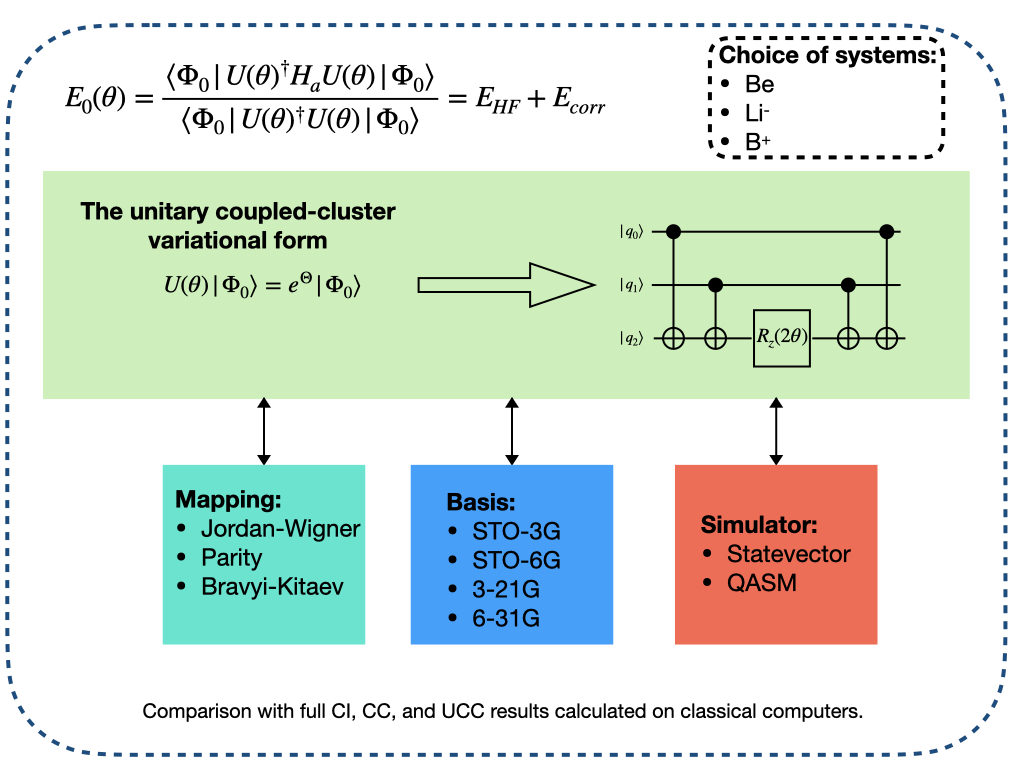} &  \includegraphics[width=9.6cm, height=9.2cm]{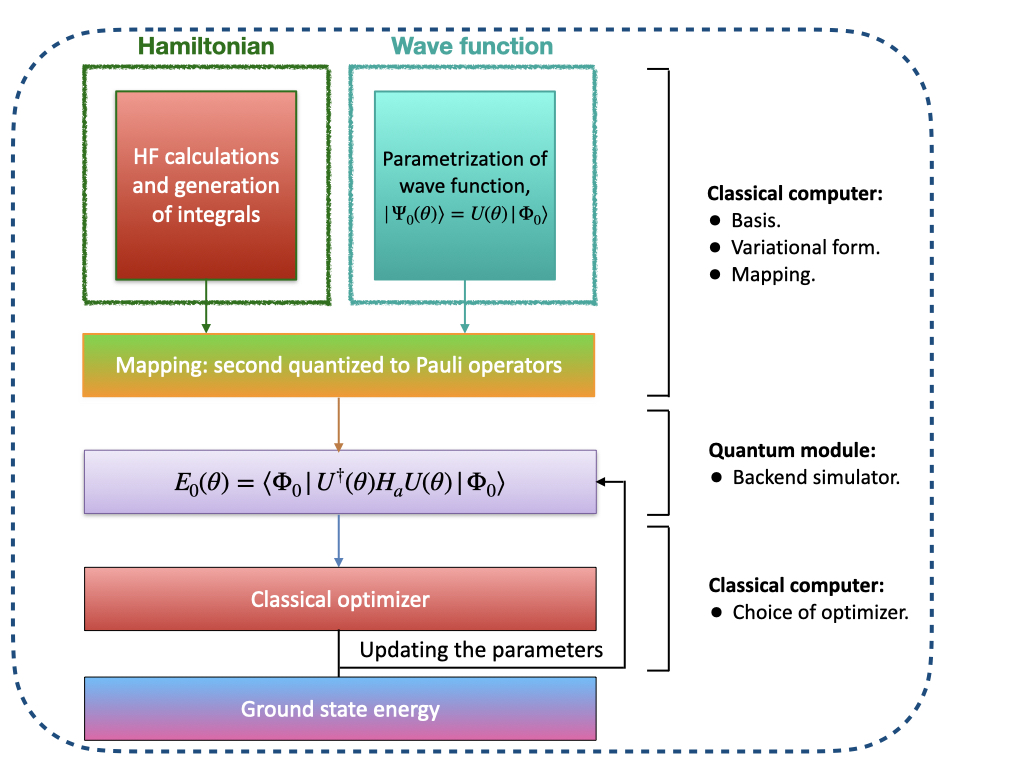} \\
    (a) & (b) \\
\end{tabular}
\caption{(a) A schematic demonstrating evaluation procedure of Eq. (\ref{engex}) using the VQE algorithm. It includes different combinations of mapping (in green), basis sets (in blue), and backend simulator (in red) to capture the correlation effects. A sample template UCC circuit is provided for the case of three qubits ($q_0$, $q_1$, and $q_2$), built out of CNOT and $R_z(2 \theta)$ gates. The figure lists the atomic systems chosen for the investigation, and it also mentions about the comparison of our VQE results with those obtained on a classical computer (ClC) by employing various many-body methods. (b) An overview of the VQE algorithm applied to electronic structure problem, which requires generating the one-body and two-body integrals of atomic Hamiltonian, expressing the wave function in parametric form and adopting quantum modules for computation. The guess parameters are then updated each time in an iterative procedure on a classical optimizer until a global minimum is reached.}
\label{fig:figure1}
\end{figure*}

Precise quantum many-body calculations in atoms and molecules are based on a rigorous treatment of the electron correlation effects. Although simulations of electronic structure have been performed using quantum algorithms, not much emphasis has always been placed on obtaining the correct correlation trends, mostly owing to the proof-of-principle nature of the calculations~\cite{ak}. Moreover, energies of a whole host of molecular systems, such as $\mathrm{H_2O}$ \cite{ag1}, $\mathrm{H_2}$ \cite{lanyon,ag3,ak} (also see Ref. \cite{exc} for an excited state treatment using an extended version of VQE), $\mathrm{HeH^+}$ \cite{peruzzo,hehplus}, LiH, BeH$_2$ \cite{ak}, and $\mathrm{H_4}$ \cite{h4}, have been calculated, but atomic systems have received little attention, except for one work on H$^-$, which is a relatively simple system~\cite{panigrahi}, in spite of finding many applications~\cite{atom1, atom2, atom3, atom4, atom5, atom6, atom7}. Atomic systems, in our view, merit separate study, since they have been and are still being used in testing new physics, such as parity violation and electric dipole moments of quarks. Atoms can provide insights on a variety of physics problems such as those mentioned in the previous sentence, via many-body calculations of relevant properties. Atomic systems are still considered as the \textit{most} suitable candidates for making atomic clocks, probing nuclear structures by studying isotope shifts, investigating fundamental physics such as new physics beyond the standard model of particle physics by analysing atomic parity violation and precise values of $g_j$ factors, and many more. All these studies entail performance of high-accuracy atomic calculations (even less than 1\% level). Many of such studies are performed in heavier atomic systems, for which quantum computers will be more appropriate than classical ones, when more qubits will be available in the future. At this point, we will comment on evaluating properties other than energy using the VQE algorithm, and their importance when more qubits become available. The converged parameters from a VQE calculation are used in constructing the wave function, which is used in calculating the energies. One then evaluates a property of interest with the converged amplitudes and appropriate property integrals. These include atomic properties of interest like the hyperfine structure constants. With appropriate modifications, this approach can be used to calculate properties of interest to fundamental physics (such as probing the electric dipole moment of the electron), and properties like dipole polarizabilities for atomic clocks. Atoms cannot be viewed as subsets of molecules, in that the correlation effects and trends in a molecule and its constituent atoms can be quite dissimilar. Atomic systems have shown to display their own unique features in this regard, in particular the conservation of orbital angular momentum in these systems~\cite{Lindgren}. Moreover, atoms are better platforms than molecules for testing the dependence of properties on the number of qubits, since the latter is composed of two or more atoms, and hence the required number of qubits, in general, grow much faster when one goes from lighter to heavier systems. In this work, we conduct a study on carefully chosen atomic systems, in which we strive to understand the precision with which the all-important electron correlation effects are captured by quantum simulations using the VQE algorithm. Specifically, within a given basis set, we check with different combinations of fermionic to qubit operator mapping and backend simulator if our quantum simulation results lie within the neighborhood of the best possible result within that basis, thus setting a measure for the precision of our results. In addition, we compare our results with those obtained from a traditional computation by using several many-body methods.  Since this is the first study of this kind on atomic systems, we strongly believe that the results and conclusions from this work will pave the way for further works on atoms. This will be a new and refreshing addition to the otherwise common approach of going up in the length scale, from diatomics to polyatomics and aimed at eventually moving to drug design on complex molecules etc, to moving in the opposite direction to atomic systems, which we reiterate has somehow received little attention. 

On physical grounds, many-body effects are expected to behave differently in ions and neutral atoms of isoelectronic systems. Among them, electron correlation effects in the negative ions are vastly different \cite{kello,bksahoo} owing to the short-range potentials that bind the outer valence electron in these ions \cite{andersen}. Negative ions find several applications, which is evident from the sheer volume of literature on them~\cite{andersen,massay,dudinikov}. Also, atomic calculations from earlier works have shown that electron correlation effects in the alkaline earth-metal atoms are very prominent due to strong repulsion between the outer two valence electrons in these atoms \cite{lim,sahoo,yashpal}. For these reasons and keeping in mind the steep cost of simulation in the NISQ era, we consider here isoelectronic neutral beryllium (Be), lithium anion (Li$^-$), and boron cation (B$^+$) as representative systems to investigate roles of electron correlation effects in the determination of their ground state energies. We also stress on the fact that the study undertaken in this work is general in nature, and should be applicable to other heavier atomic systems in higher quality basis sets, when such simulations become feasible. It is also worth adding that the systems that have been investigated in this work find many applications. For example, light systems such as Be can serve as excellent systems in probing roles of different kinds of electromagnetic interactions \cite{Beqed,Beqed2}, as well as obtaining nuclear charge radii from measurements of isotope shifts~\cite{Beisotopeshift}. Moreover, Be is a very interesting system from a many-body theoretic viewpoint, as it is well known that its ground state has a multireference character~\cite{Lindgren}. Systems such as Li$^-$ may find applications in plasma diagnostics~\cite{Liminus}. Group IIIA ions have been known to hold great promise for atomic clocks~\cite{bpd1}. Specifically, B$^+$, holds promise, since the transition of interest has an extremely long lifetime in its excited state. Moreover, because the $^{10}$B$^+$ ion's mass is closer to that of $^9$Be$^+$, there would be efficient state exchange for quantum logic detection~\cite{wineland}.

The accuracy of the calculated ground state energy of a system using a VQE algorithm depends upon several factors, including the crucial aspect of choosing a variational form, as it dictates the form of the wave function. The other elements that need special attention are the choice of mapping technique used to convert the second quantized fermionic operators describing the Hamiltonian and wave function to their spin counterparts, the backend simulator for running quantum circuits, and the classical optimizer, besides the more intuitive and traditional features such as the choice of single-particle basis in the many-body calculations. We focus extensively on the required number of shots for obtaining reliable results using Qiskit's QASM simulator backend. Employing the QASM simulator marks a significant departure from the otherwise common use of the statevector backend (for example, see Refs.~\cite{sv1,sv2,sv3,sv4}. Some works such as Ref.~\cite{qasm1} use QASM, but with much fewer shots (4096) than possibly required). While both the backends are used for the same VQE algorithm, the latter relies on matrix manipulations whereas the former is measurement-based. Our investigation is especially necessary, as it explicitly provides estimates for expected error from a measurement-based scheme. This sets the ground for future analyses where one may include noise models and error mitigation, which then would be more realistically comparable to a calculation performed on a real quantum computer. 

\begin{figure}[t]
    \centering
    \begin{tabular}{cc}
        \includegraphics[width=8.5cm,height=7cm]{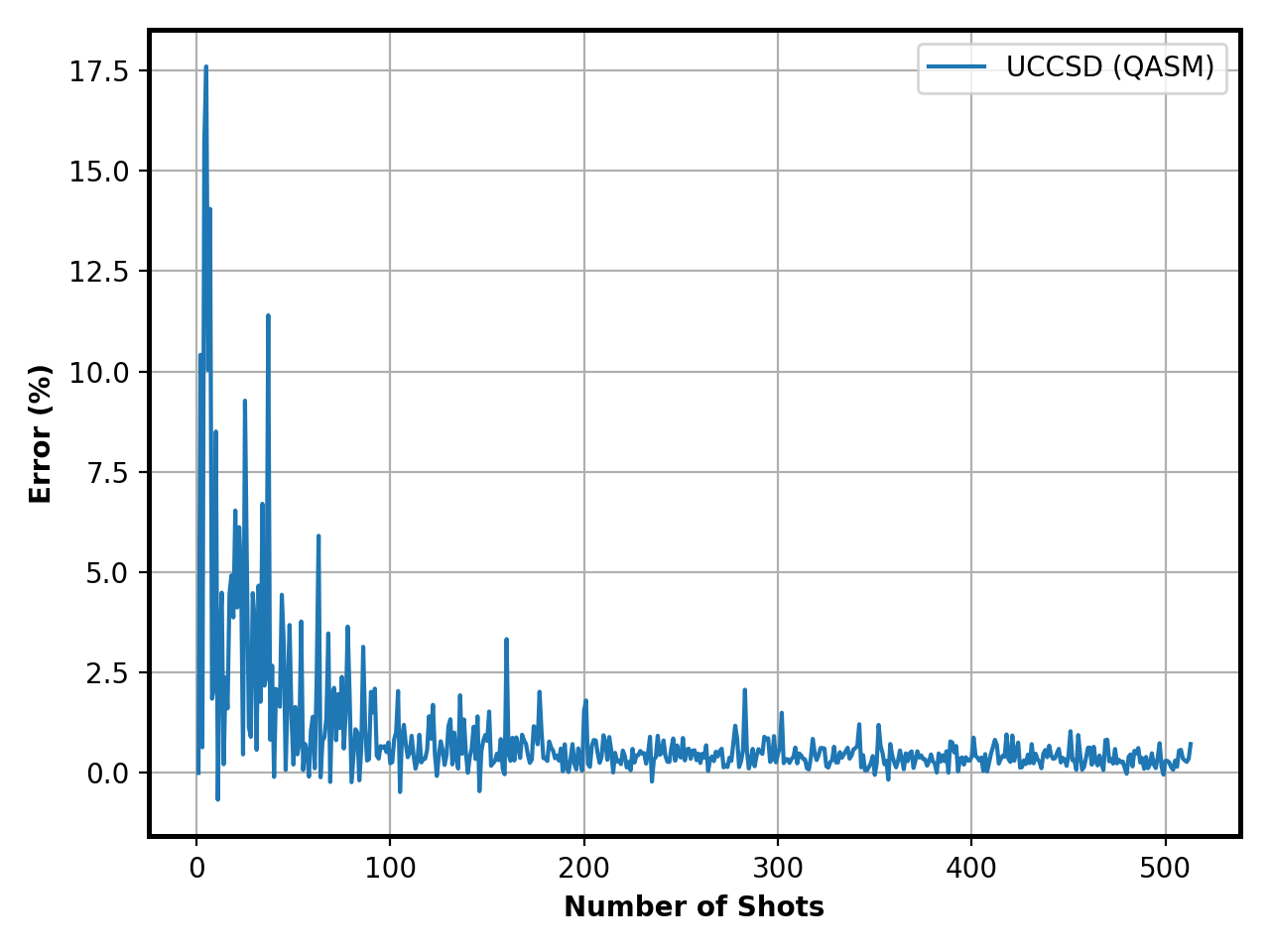}
    \end{tabular}
    \caption{Plot showing the variation in the percentage fraction error in the calculated energy using the QASM simulator with respect to the FCI value, with the number of shots up to 512, for Be in the STO-3G basis and with the JW mapping.}
    \label{fig:shots}
\end{figure}

\section{Theory}\label{theory}

\subsection{The VQE Algorithm}

The ground state energy functional of an atomic system within the classical-quantum hybrid VQE algorithm is given as 
\begin{eqnarray}
E_0(\theta) &=& \frac{\langle \Psi_0(\theta)\arrowvert H_a \arrowvert \Psi_0(\theta)\rangle}{\langle \Psi_0(\theta) \arrowvert \Psi_0(\theta)\rangle} \nonumber \\
&=&\langle \Phi_0 \arrowvert U^\dag(\theta) H_a U(\theta) \arrowvert \Phi_0\rangle, 
\label{engex}
\end{eqnarray}
where $\arrowvert \Psi_0 (\theta) \rangle$ is the trial wave function and $H_a$ is the atomic Hamiltonian. The former is parametrized as $\arrowvert \Psi_0 (\theta) \rangle = U(\theta) \arrowvert \Phi_0 \rangle$, where the unitary operator, $U(\theta)$, depends on a set of arbitrary parameters, denoted collectively as $\theta$, and $\arrowvert \Phi_0 \rangle$ is a suitable initial many-body wave function. For the choice of $U(\theta)$, that is, the {\it ans\"{a}tz}, we adopt the unitary coupled-cluster (UCC) variational form. The initial state is constructed by employing the Hartree-Fock (HF) approximation. In this framework, the total Hamiltonian is expressed as $H_a=H_0 + V_{es}$, constructed out of the effective one-body HF Hamiltonian $H_0$ and the residual interaction term $V_{es}$ \cite{Lindgren}. Therefore, the total energy is computed as $E_0=E_{HF}+E_{corr}$ with the HF energy $E_{HF}=\langle \Phi_0 | H_a | \Phi_0 \rangle$ and correlation energy $E_{corr}$ arising from the residual interaction $V_{es}$. The UCC theory accounts for these correlation effects via an exponential {\it ans\"atz} acting on the HF wave function, with the exponent expressed as the sum of excitation operators. Further, in this {\it ans\"{a}tz}, the Trotterization \cite{trotter} procedure is used to decompose $U(\theta)$ into smaller operators to implement it efficiently in a quantum circuit. Throughout, we work with $\arrowvert \Psi_0 (\theta) \rangle$ in its second quantized form, where the relevant mathematical structures are recast in the language of creation and annihilation operators. In Eq. (\ref{engex}), the atomic Hamiltonian, $H_a$, too is expressed in the second quantized form as 
\begin{equation}\label{hamiltonian}
    H_a = \sum_{pq}^D h_{pq} a^{\dagger}_p a_q + \frac{1}{2}\sum_{pqrs}^D h_{pqrs} a^{\dagger}_p a^{\dagger}_q a_r a_s, 
\end{equation}
where $h_{pq}$ and $h_{pqrs}$ denote one-body and two-body integrals of $H_a$, respectively, $D$ refers to the number of spin-orbitals from the chosen single particle basis, and the notations $\{p,q,r,s\}$ denote general atomic orbitals. 

To compute atomic energies in the framework of quantum simulation, one needs to express the second quantized fermionic operators that occur in both the wave function as well as the Hamiltonian as spin operators that contain a sequence of unitary operations. We use three such mapping techniques, namely the Jordan-Wigner (JW), Parity (PAR), and the Bravyi-Kitaev (BK) transformations.  
In the JW transformation \cite{jwt}, one works in the occupation number basis. For a given site, $k$, the creation and annihilation operators, $a_k$ and $a_k^\dag$, are related to their corresponding gate structures, $A_k$ and $A_k^\dag$, via the JW transformation, by 
\begin{equation}\label{Eq:jw1}
{A_k}^\dagger = \otimes_{j=1}^{k-1} Z_j \otimes {Q_k}^+ \otimes_{j= k+1}^D I_j, 
\end{equation}
and 
\begin{equation}\label{Eq:jw2}
{A_k} = \otimes_{j=1}^{k-1} Z_j \otimes {Q_k^-} \otimes_{j= k+1}^D I_j,   
\end{equation}
where $Q_k^+$ and $Q_k^-$ are given by 
\begin{equation}\label{q1}
Q_k^+ = \frac{X_k + \dot{\iota} Y_k}{2},  
\end{equation}
and 
\begin{equation}\label{q2}
Q_k^- = \frac{X_k - \dot{\iota} Y_k}{2}, 
\end{equation}
$D$ refers to the total number of qubits for the considered system,
and $X$, $Y$, and $Z$ are the Pauli operators/gates. The string $\otimes_{j=1}^{k-1} Z_j$ in Eqs. (\ref{Eq:jw1}) and (\ref{Eq:jw2}) ensures that in the spin operator-transformed version, the required phase change that occurs when a creation or annihilation operator acts on an arbitrary Fock state is accounted for. In the case of atoms, this would correspond to $D$ spin-orbitals, with $N$ electrons, or in other words, $N$ occupied and $(D-N)$ unoccupied spin-orbitals. Note that for the description of a given $A_k$, one needs to take into account a tensor product of $Z$ gates that contains $(k-1)$ terms. Therefore, when one constructs terms such as $A_i^\dag A_j + A_j^\dag A_i$, which one normally encounters in atomic calculations, the resulting term has tensor product of Pauli gates from index $i$ through $j$, with \textit{all} the in-between indices occurring in steps of one. Thus, the number of qubit operations for JW transformation scales as $\mathcal{O}(D)$. In case of PAR transformation, the $k^{th}$ qubit stores the parity of all the spin-orbitals up to $k$. Thus, PAR transformation is known to work in the parity basis. The PAR mapping uses two-qubit reduction arising from the $Z_2$ symmetry, thereby reducing the number of required qubits by two. We note that the PAR transformation too scales as $\mathcal{O}(D)$. The BK transformation works in the BK basis, which finds a golden mean by borrowing from both of above mentioned approaches, and leads to lesser number of required gates. The number of qubits scales as $\mathcal{O}(log D)$. We have not provided the relevant equations for the PAR and the BK transformations as we have for the JW mapping, and an interested reader can find a comprehensive discussion on all these three transformations in Refs.~\cite{jwt,z2}. 

The circuits thus constructed by starting from Eq. (\ref{engex}) are evaluated with an initial set of guess parameters in an appropriate backend simulator (either statevector or qiskit's QASM backend), and the energy is obtained. The statevector simulator executes the set of circuits associated with a system without measurements or shots, given an input state vector. The QASM simulator mimics an ideal quantum computer, in that it gives probabilistic outcomes as counts for each of the states, after multiple shots. After evaluating Eq. (\ref{engex}), we pass the energy to an optimizer, which runs on a ClC. This module uses an optimization algorithm, and minimizes the energy, obtained from the previous step of the VQE algorithm, with respect to the parameters. Once the new parameters are obtained, they are fed back as inputs to the quantum circuit from the previous step. This process is repeated until the energy is minimized. The energy thus obtained is guaranteed to be an upper bound to the true ground state energy. 

\subsection{Many-body methods}\label{sec3}

The ground state energy, $E_0$, with the exact wave function $\arrowvert \Psi_0 \rangle$ of an atomic system can be determined by
\begin{eqnarray}\label{eq:0}
E_0 = \frac{\langle \Psi_0 \arrowvert {H_a} \arrowvert \Psi_0 \rangle}{\langle \Psi_0 \arrowvert \Psi_0 \rangle} .
\end{eqnarray}
The full configuration interaction (FCI) method can be employed to determine on $\arrowvert \Psi_0 \rangle$ exactly for a given basis set by expressing as 
\begin{eqnarray}
|\Psi_0 \rangle = C_0 |\Phi_0 \rangle + C_I |\Phi_I \rangle + C_{II} |\Phi_{II} \rangle + \cdots + C_N |\Phi_N \rangle, \ \ \
\end{eqnarray}
where $\{C\}$s are the expansion coefficients with the Slater determinants $\{|\Phi\rangle\}$s generated by exciting the HF wave function $\arrowvert \Phi_0 \rangle$. Due to extremely steep computational cost, truncated configuration interaction (CI) method is usually considered in the multi-electron systems for practical scenarios. At a given level of truncation, the coupled-cluster (CC) theory accounts for electron correlation effects more rigorously and satisfies size-consistency and size-extensivity characteristics in contrast to the CI method, thereby earning the title of the gold standard of electronic structure calculations \cite{Bartlett}. In the CC theory {\it ans\"atz}, $\arrowvert \Psi_0 \rangle$ yields the exponential form as (e.g. see Refs. \cite{Bartlett,springer})
\begin{eqnarray}
\arrowvert \Psi_0 \rangle = e^{T} |\Phi_0 \rangle,
\end{eqnarray}
where $T=T_1 + T_2 + ... +T_N$ is a sum of excitation operators generating particle-hole excitations with the level denoted by subscript. The amplitudes of these operators are obtained on ClC by solving the following equation
\begin{eqnarray}
\langle \Phi_0^K | {\cal H}_a | \Phi_0 \rangle = 0,
\end{eqnarray}
where ${\cal H}_a \equiv e^{-T} H_a e^T = \left ( H_a e^T \right )_l$ with subscript $l$ representing the linked terms~\cite{Bartlett}. The $K$ on the right hand side of the above equation denotes the $K^{th}$ excitation out of the HF state. For example, in the CCSD method, there are two amplitude equations, one with $K$ denoting single excitations, and the other with $K$ specifying double excitations. Once the amplitudes associated with the $T$ operators (t-amplitudes) are obtained, the energy of the system is calculated  by 
\begin{eqnarray}
E_0 &=& \langle \Phi_0 | {\cal H}_a  | \Phi_0 \rangle \nonumber \\ 
   &\ne & \langle \Phi_0 | e^{T^{\dagger}} H_a e^T \arrowvert \Phi_0 \rangle,
  \label{eqeng}  
\end{eqnarray}
where the inequality sign indicates that the energy expression given by Eq. (\ref{eqeng}) is not variational in an approximated CC theory owing to non-hermitian property of ${\cal H}_a$, but it terminates naturally. 

It is desirable to work with unitary operators in the framework of quantum computation/simulation. For this purpose, we take recourse to the UCC theory over the CC theory. In the approximated UCC theory framework~\cite{ucc}, $U(\theta)  = e^{\Theta(\theta)}$ with $\Theta(\theta)=T - T^{\dagger}$ such that t-amplitudes are used as $\theta$. One can immediately see from the above equation that the UCC operator involves not only the excitation operator ${T}$ but also the de-excitation operator $T^{\dagger}$. The energy expression follows
\begin{eqnarray}
E_0 &=& \langle \Phi_0 | e^{\Theta^{\dagger}} H_a e^{\Theta} \arrowvert \Phi_0 \rangle  
    = \langle \Phi_0 | e^{-\Theta} H_a e^{\Theta} \arrowvert \Phi_0 \rangle  \nonumber \\
    &=& \langle \Phi_0 | e^{T^{\dagger} -T} H_a e^{T - T^{\dagger}} \arrowvert \Phi_0 \rangle. \label{ucc_expression}
\end{eqnarray}
Unlike in the traditional version of the CC method, $e^{T^{\dagger} -T} H_a e^{T- T^{\dagger}}$ does not terminate naturally in the above equation, but it guarantees that the energy thus calculated obeys the variational principle. Owing to the non-terminating form,  Eq. (\ref{ucc_expression}) cannot be evaluated efficiently in the traditional UCC method on a ClC without resorting to brute-force termination of the expression. However, this issue is circumvented on a quantum computer/simulator. 

To carry out the analysis conveniently, we have used the approximated CC theory to singles (S) and doubles (D) approximation (CCSD method). The singles and and doubles approximated UCC theory is henceforth mentioned as the UCCSD method. The S and D level excitations are denoted by subscripts $1$ and $2$ respectively, and these operators are defined in the second-quantized form as
\begin{eqnarray}
T &\approx& T_1 + T_2 = \sum_{ia} \tau_{ia} a^{\dagger}_i a_a + \frac{1}{4}\sum_{ijab} \tau_{ijab} a^{\dagger}_i a^{\dagger}_j a_a a_b , \ \ \
\label{tamp}
\end{eqnarray}
where $\tau$s are the t-amplitudes, and notations $\{a,b\}$ and $\{i,j\}$ denote occupied and virtual orbitals, respectively. 

\section{Methodology}\label{methodology}

We carried out the FCI and CCSD calculations using PySCF \cite{pyscf}, while the UCCSD computations were performed using the OpenFermion-PySCF~\cite{McClean} program. The one-body and two-body integrals, as can be seen from Eq. (\ref{hamiltonian}), are the main ingredients from a ClC to carry out many-body calculations on a quantum simulator. These integrals are obtained from the PySCF program \cite{pyscf}. In this program, Gaussian type orbitals~\cite{Boys}, specifically contracted versions of the minimal STO-3G and STO-6G basis~\cite{sto}, as well as Pople's 3-21G basis and 6-31G basis~\cite{Pople}, are employed. Since the number of qubits required for the computations is equal to the number of spin-orbitals (which is in turn decided by the choice of single-particle basis set), the qubit requirement for Be, Li$^-$, and B$^+$ is 10 for the STO-3G and STO-6G basis sets, while it is 18 for the 3-21G and 6-31G basis sets. We stress that we carry out all-electron calculations, that is, we do not freeze any of the occupied spin-orbitals. This factor, in combination with the chemically (and/ or physically motivated) UCCSD variational form, leads to the computations becoming expensive. As an example, Be in the 6-31G basis, which is a 18 qubit computation, demands for about 1900 gates even with the ‘heavy’ $RYRZ$ with full entanglement strategy hardware efficient ans\"{a}tz, but UCCSD demands for about 32000 gates. This scaling makes computations with more qubits challenging. In all of our calculations, we set the initial guess parameters for the variational form to zero. Also, we fixed the Trotter number to be one. We used a gradient-free approach, the COBYLA (Constrained Optimization BY Linear Approximation) optimizer, which is commonly used in literature~\cite{cobyla,h4, cb1, cb2}. For an optimization problem with $N$ design variables, a simplex of $N+1$ vertices is constructed. Hereafter, a linear polynomial approximation is used as an interpolation of the objective function and the inequality constraints of the problem. The algorithm controls the size of the trust region (simplex) and decreases it until a convergence is reached. The convergence for COBYLA optimizer is slower than the gradient based methods as it requires higher number of function evaluations to reach the optimum value. However, stability comes as a notable feature for this algorithm along with lesser number of parameters to be tuned for performing optimization \cite{cobpar}. We used the qiskit 0.15.0 package \cite{qiskit} to carry out quantum simulations using the VQE algorithm.

We have depicted the important steps followed in the current work for better understanding of its objective in Fig.~\ref{fig:figure1}(a), while the general features of the VQE algorithm adopted in the present work, as well as its structure is encapsulated in Fig.~\ref{fig:figure1}(b). The qubit mapping of UCC operators into circuit form includes a rotation gate, $R_z(2 \theta)$, encased within a staircase structure constructed out of two-qubit CNOT-gates. Fig.~\ref{fig:figure1}(a) shows an example with 3 qubits, $q_0$, $q_1$, and $q_2$, where the circuit represents $U(\theta) = e^{-i \theta Z_0Z_1 Z_2}$ with $Z_i$ referring to the $i^{th}$ Pauli $Z$-gate. For the cases where the exponent contains the $X$ or $Y$ Pauli operators, the basis is rotated with the appropriate single-qubit rotation gates.

In the Results section, we show the dependence of the calculated ground state energies of Be, Li$^-$, and B$^+$ using the VQE algorithm on combinations of different mappings and simulators, within a basis set. For the larger 3-21G and the 6-31G bases, we only provide results obtained with the statevector simulator. We also give the HF, CCSD, UCCSD, and FCI calculations, obtained with a ClC, for comparison. Explicitly giving the HF energy allows us to visually check for the correlation effects captured by a VQE calculation for a given combination of basis, mapping, and backend. 

\section{Results and Discussion}\label{sec2}

We first present an outline of the contents of this section in order to make it easier to follow the organization of the results. We first discuss the analysis of the required number of shots for QASM calculations, with the corresponding figures being Figs. 2 and 3. This is followed by a brief analysis of errors due to Trotter number. We then move to the main results pertaining to the calculations with the STO-3G basis, followed by those from the STO-6G basis. We then examine the results of our  calculations with the 3-21G and the 6-31G bases, and provide additional comments on possible discrepancies due to optimizers,  STO-6G vs 3-21G bases, and comparison of results across smaller (STO family) and larger (split-valence, that is, 3-21G and 6-31G) bases. 

\begin{figure}[t]
    \centering
    \includegraphics[width=8.5cm, height=7cm]{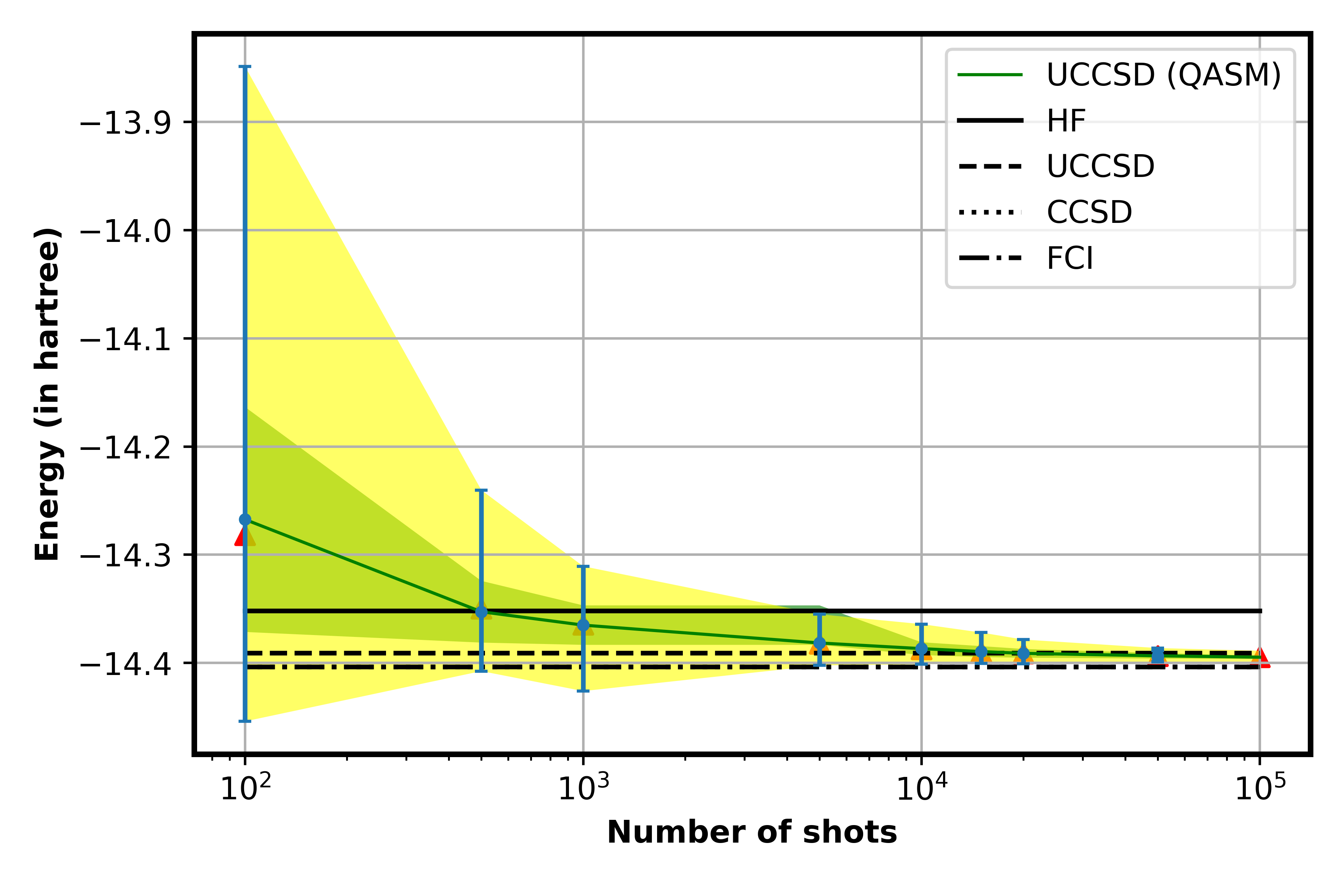} 
    \caption{Analysis of UCCSD energy from QASM (UCCSD(QASM)) versus the number of shots, for Be in STO-3G basis and with JW mapping. The results obtained using the QASM backend are also compared with the values obtained using the FCI, CCSD, and UCCSD methods in a ClC. Each data point (circle) represents the mean of 160 runs for a given number of shots, and is accompanied by two error bars, with the band in yellow quoting the range (maximum - minimum), while the green band denotes the standard deviation. The data points marked with a triangle refer to the bootstrapped mean. }
    \label{fig:25k_shots}
\end{figure}

\setlength{\tabcolsep}{8pt}
\begin{table*}[t]
\caption{A quantitative analysis of the ground state energies (in hartree) of  Be, Li$^-$, and B$^+$ computed using the VQE algorithm in the STO-3G basis and adopting the UCCSD ans\"{a}tz, with different combinations of simulators and fermion to qubit mapping techniques. The results are compared with values obtained using various methods in a ClC. Next to the final values, we provide the correlation energy for that combination in brackets. The percentage fraction difference with respect to FCI results is denoted as `$\Delta$ in $\%$'.}
    \centering
    \setlength{\tabcolsep}{0mm}
    \renewcommand{\arraystretch}{1.2} 
    \begin{tabular}{c<{\hspace{7mm}}c<{\hspace{5mm}}c<{\hspace{5mm}}c<{\hspace{5mm}}c}
          \hline \hline
          Mapping & Method &  Be &Li$^-$ &  B$^+$ \\
          \hline
           \addlinespace[0.1cm] \\
   & HF  & $-$14.351880 & $-$7.213273  & $-$23.948470    \\ 
  ClC    & FCI     & $-$14.403655($-$0.051775) & $-$7.253791($-$0.040518) &$-$24.009814($-$0.061344)  \\
   &  CCSD  & $-$14.403651($-$0.051771)  & $-$7.253786($-$0.040513)  & $-$24.009811($-$0.061341)  \\
    & UCCSD   & $-$14.391028($-$0.039148) & $-$7.244008($-$0.030735) & $-$23.994757($-$0.046287)  \\
    \\
  & UQ   & $-$14.388109($-$0.036229)  & $-$7.244270($-$0.030997)   & $-$24.002041($-$0.053571)   \\
   JW  &  ($\Delta$ in \%)  & ($-0.108$) & ($-0.131$) &  ($-0.032$) \\
  & US   & $-$14.403490($-$0.05161) & $-$7.253682($-$0.040409)  & $-$24.009652($-$0.061182)  \\
   & ($\Delta$ in \%)  & ($-0.001$) & ($-0.001$)&  ($-0.001$) \\
   \\
    & UQ    & $-$14.394762($-$0.042882) & $-$7.243156($-$0.029883)  & $-$23.992675($-$0.044205)   \\
  PAR  & ($\Delta$ in \%)  & ($-0.062$) & ($-0.146$) & ($-0.071$) \\
   & US    & $-$14.403446($-$0.051566) & $-$7.253611($-$0.040338)   & $-$24.009631($-$0.061161)    \\
   & ($\Delta$ in \%)  & ($-0.001$) & ($-0.002$) & ($-0.001$) \\
   \\
   & UQ    & $-$14.392365($-$0.040485)  &    $-$7.243775($-$0.030502)  & $-$23.998311($-$0.049841)   \\
   BK    & ($\Delta$ in \%) & ($-0.078$)  & ($-0.138$) & ($-0.048$) \\
   & US   & $-$14.403539($-$0.051659)   &  $-$7.253681($-$0.040408) & $-$24.009500($-$0.06103)  \\
    & ($\Delta$ in \%)  & ($-0.001$) & ($-0.001$) & ($-0.001$) \\ \\
        \hline \hline
    \end{tabular}
    \label{tab:s3g}
\end{table*}

\setlength{\tabcolsep}{8pt}
\begin{table*}[t]
\caption{The ground state energies of Be, Li$^-$, and B$^+$ obtained using the STO-6G basis. All notations are the same as in Table \ref{tab:s3g}. 
}
    \centering
    \setlength{\tabcolsep}{0mm}
    \renewcommand{\arraystretch}{1.2} 
    \begin{tabular}{c<{\hspace{7mm}}c<{\hspace{5mm}}c<{\hspace{5mm}}c<{\hspace{5mm}}c}
    \hline \hline
          Mapping & Method &  Be &Li$^-$ &  B$^+$   \\
          \hline \\
        \addlinespace[0.1cm]
        & HF   & $-$14.503361 & $-$7.295246 & $-$24.190562   \\
        ClC & FCI & $-$14.556088 ($-$0.052727) & $-$7.336640 ($-$0.041394) & $-$24.252889 ($-$0.062327)   \\
         &  CCSD    & $-$14.556083 ($-$0.052722) & $-$7.336635 ($-$0.041389) & $-$24.252884 ($-$0.062322)  \\ 
        & UCCSD  & $-$14.543257 ($-$0.039896)  & $-$7.326677 ($-$0.031431)  & $-$24.237615 ( $-$0.047053) \\ \\
       \addlinespace[0.1cm]
       & UQ     & $-$14.544091 ($-$0.04073) &  $-$7.326529 ($-$0.031283) & $-$24.227757 ($-$0.037195)  \\
       JW & ($\Delta$ in \%)  &  ($-0.082$) & ($-0.138$) & ($-0.055$) \\ 
      & US     &  $-$14.555940 ($-$0.052579)  & $-$7.336485 ($-$0.041239)  & $-$24.252614 ($-$0.062052)   \\
        & ($\Delta$ in \%)  & ($-0.001$)  & ($-0.002$) & ($-0.001$) \\ \\
      \addlinespace[0.1cm]
       & UQ      & $-$14.541160 ($-$0.037799) & $-$7.321997 ($-$0.026751) & $-$24.222150 ($-$0.031588) \\
      PAR  & ($\Delta$ in \%)   & ($-0.103$) & ($-0.199$) & ($-0.127$) \\
        & US    & $-$14.555943 ($-$0.052582) & $-$7.336510 ($-$0.041264) & $-$24.252623 ($-$0.062061)    \\
         & ($\Delta$ in \%)    & ($-0.001$) & ($-0.002$) & ($-0.001$) \\ \\
       \addlinespace[0.1cm]
       & UQ    & $-$14.548048 ($-$0.044687)  & $-$7.321989 ($-$0.026743) & $-$24.241578 ($-$0.051016)   \\
        BK & ($\Delta$ in \%)  & ($-0.055$)  & ($-0.199$) & ($-0.047$) \\
     & US    & $-$14.555848($-$0.052487)  & $-$7.336462 ($-$0.041216)  & $-$24.252669 ($-$0.062107)  \\
    & ($\Delta$ in \%)  & ($-0.002$)  & ($-0.002$) & ($-0.001$) \\ \\
        \hline \hline
        \end{tabular}
        \label{tab:s6g}
\end{table*}

\begin{table*}[t]
\caption{The table presents the energies for Be, Li$^-$,  and B$^+$ in the 3-21G basis. The same notations as in Table \ref{tab:s3g} are adopted here.}
    \centering
    \setlength{\tabcolsep}{0mm}
    \renewcommand{\arraystretch}{1.2} 
    \begin{tabular}{c<{\hspace{7mm}}c<{\hspace{5mm}}c<{\hspace{5mm}}c<{\hspace{5mm}}c}
    \hline \hline
          Mapping & Method &  Be &Li$^-$ &  B$^+$   \\
          \hline \\
        \addlinespace[0.1cm]
   & HF & $-$14.486820  & $-$7.366760 & $-$24.096376   \\ 
    ClC   & FCI     &  $-$14.531444 ($-$0.044624) & $-$7.397779 ($-$0.031019) & $-$24.153344 ($-$0.056968)  \\
      &  CCSD    & $-$14.531416 ($-$0.044596) & $-$7.397757 ($-$0.030997) & $-$24.153311 ($-$0.056935) \\
     & UCCSD   & $-$14.512130 ($-$0.02531) & $-$7.383818 ( $-$0.017058) & $-$24.131129 ($-$0.034753) \\ \\
        \addlinespace[0.1cm]
     JW & US  &  $-$14.513922 ( $-$0.027102)   & $-$7.385692 ($-$0.018932) & $-$24.138757 ($-$0.042381)   \\
       & ($\Delta$ in \%) & ($-0.121$)  & ($-0.163$) & ($-0.059$) \\ \\
        \addlinespace[0.1cm]
        PAR & US  &  $-$14.516600 ($-$0.02978)   & $-$7.387247 ($-$0.020487)  & $-$24.139378($-$0.043002)  \\
      & ($\Delta$ in \%) & ($-0.102$)  & ($-0.142$) & ($-0.058$) \\
         \addlinespace[0.1cm]
      BK  & US   &  $-$14.519369 ($-$0.032549)  & $-$7.386396 ($-$0.019636) & $-$24.139013 ($-$0.042637)  \\
     & ($\Delta$ in \%) & ($-0.083$)  & ($-0.154$) & ($-0.059$) \\ \\
        \hline \hline
        \end{tabular}
        \label{tab:321g}
\end{table*}

\begin{table*}[t]
     \caption{Using the same notations as in Table \ref{tab:s3g}, the ground state and correlation energies for Be, Li$^-$, and B$^+$ in the 6-31G basis are given.}
    \centering
    \setlength{\tabcolsep}{0mm}
    \renewcommand{\arraystretch}{1.2} 
    \begin{tabular}{c<{\hspace{7mm}}c<{\hspace{5mm}}c<{\hspace{5mm}}c<{\hspace{5mm}}c}
    \hline \hline
          Mapping & Method &  Be &Li$^-$ &  B$^+$   \\
          \hline \\
        \addlinespace[0.1cm]
   & HF  & $-$14.566764 & $-$7.405387 & $-$24.234041   \\ 
    ClC & FCI   & $-$14.613545 ($-$0.046781)  & $-$7.438753 ($-$0.033366) & $-$24.293125 ($-$0.059084)  \\
     &  CCSD   & $-$14.613518 ($-$0.046754)   & $-$7.438739 ($-$0.033352) & $-$24.293096 ($-$0.059055) \\
    & UCCSD   & $-$14.593071 ($-$0.026307) & $-$7.423171 ($-$0.017784) & $-$24.269635 ($-$0.035594) \\ \\
        \addlinespace[0.1cm]
      JW & US    &  $-$14.601323 ($-$0.034559)  & $-$7.426886 ($-$0.021499) & $-$24.279715 ($-$0.045674)   \\
   & ($\Delta$ in \%)  & ($-0.083$) & ($-0.159$) & ($-0.055$) \\
       \addlinespace[0.1cm]
   PAR & US  & $-$14.597296 (-0.111)   & $-$7.425017 (-0.185) & $-$24.278157 (-0.061)   \\
     &  ($\Delta$ in \%)  & ($-$0.030532) & ($-$0.01963) & ($-$0.044116) \\ \\
        \addlinespace[0.1cm]
    BK  & US   & $-$14.597296 ($-$0.030532) & $-$7.423154 ($-$0.017767)  & $-$24.277312 ($-$0.043271)    \\
    &  ($\Delta$ in \%)  & ($-0.111$) & ($-0.209$) & ($-0.651$) \\ \\
        \hline \hline
     \end{tabular}
    \label{tab:631g}
\end{table*}

\begin{figure*}[t]
    \centering
    \setlength{\tabcolsep}{1mm}
        \begin{tabular}{ccc}
            \includegraphics[width=60mm]{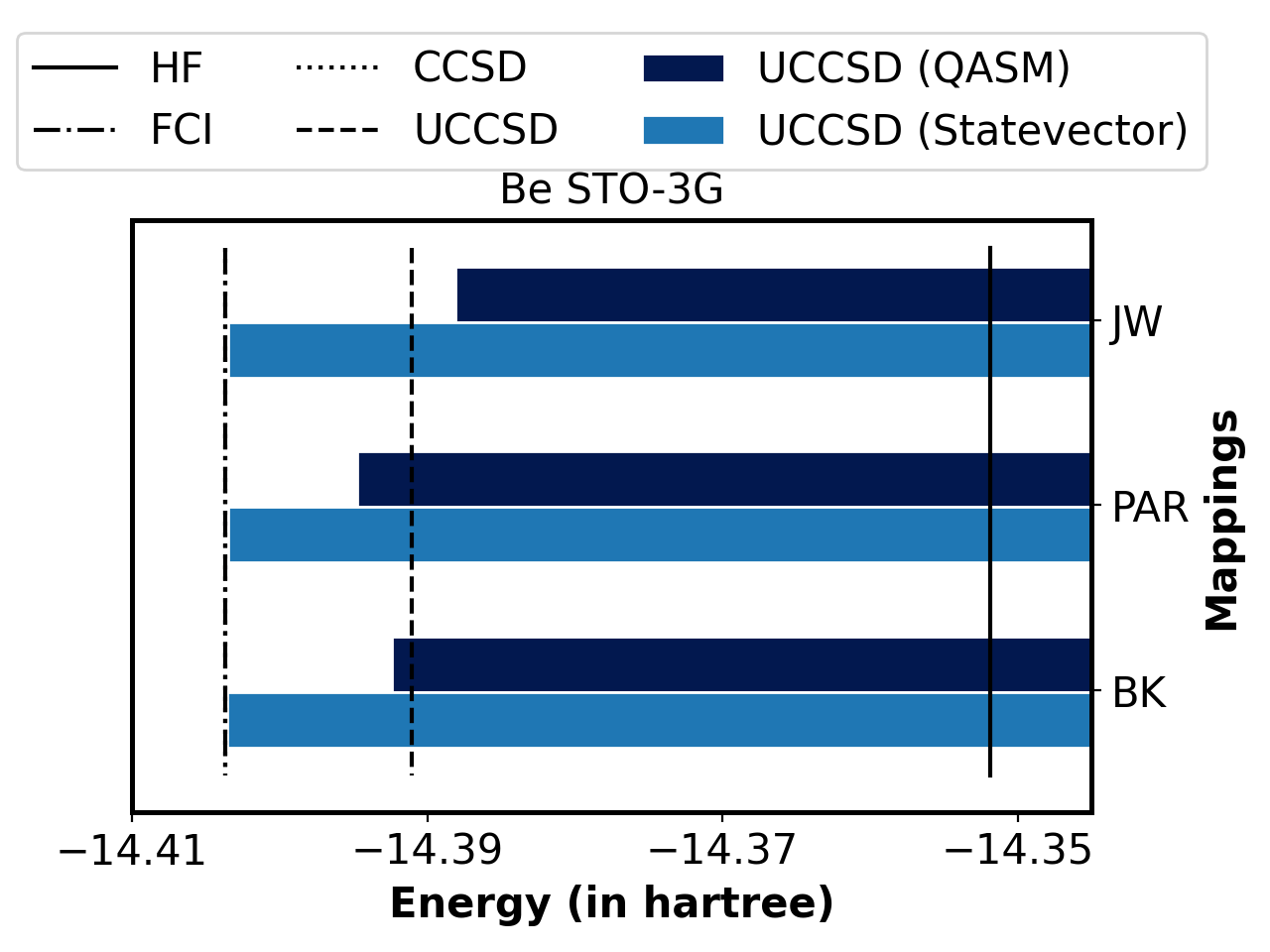} & \includegraphics[width=60mm]{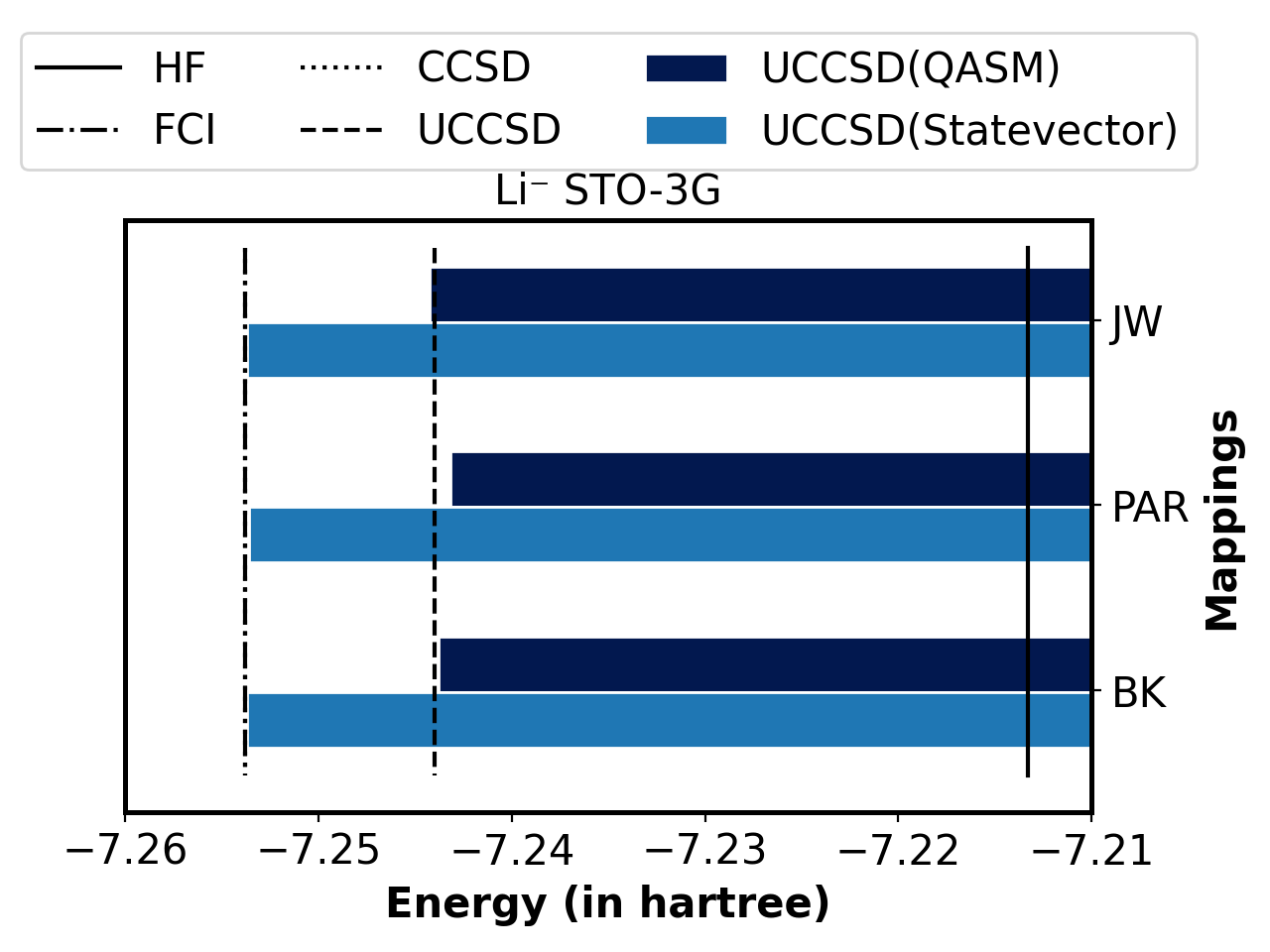} & \includegraphics[width=60mm]{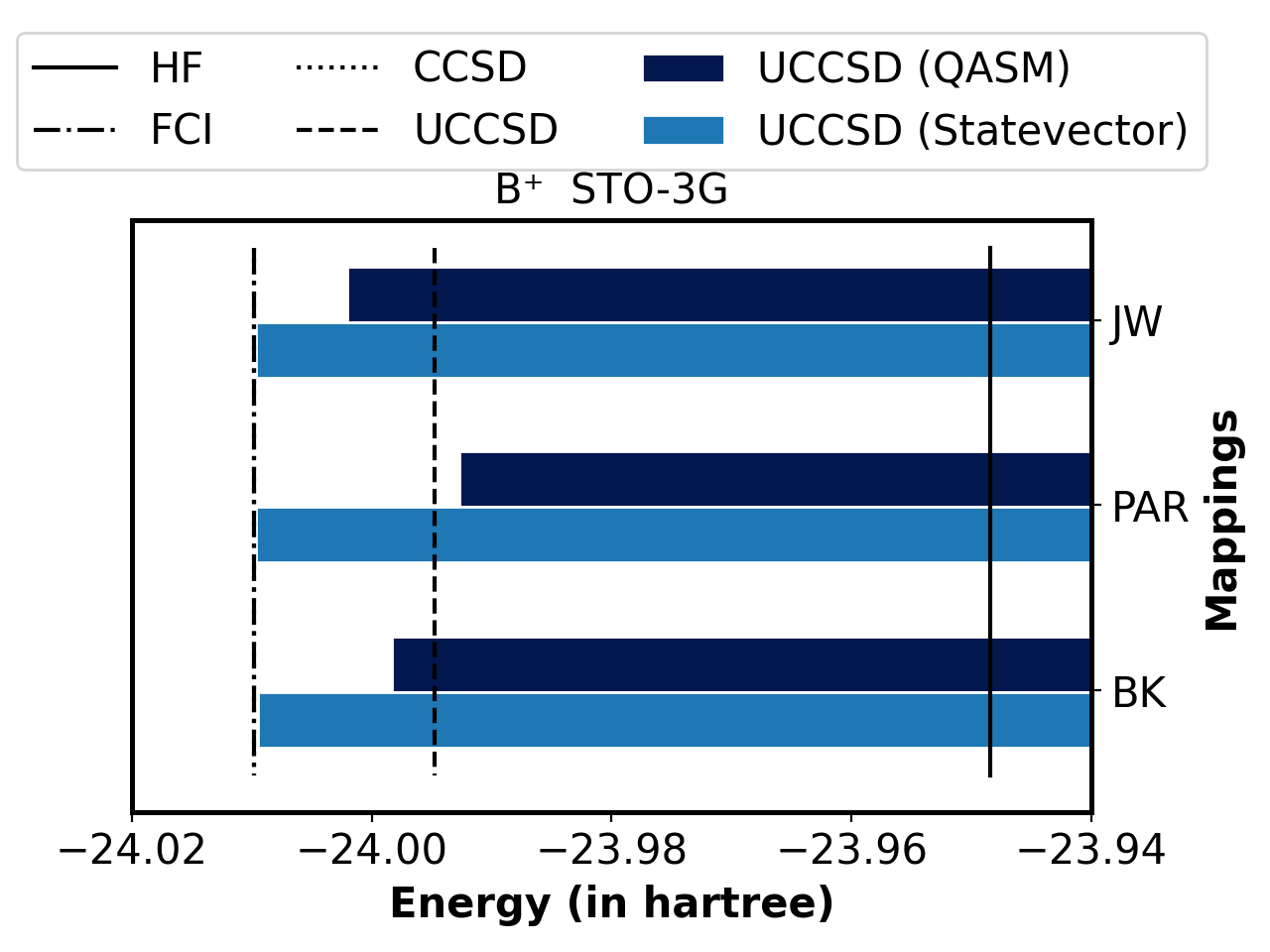} \\
            \textbf{(a)}& \textbf{(b)} & \textbf{(c)}\\
        \end{tabular}
    \caption{Graphical illustration of results for the Be, Li$^-$, and B$^+$ in STO-3G basis obtained using the VQE algorithm. The figure serves to compare the impact of different combinations of fermion to qubit mapping techniques, namely JW, PAR and BK transformations, as well as backend simulators (statevector and QASM). The dark blue bars indicate the energies obtained on a QASM simulator, while the bars in light blue specify the energies computed using a statevector simulator. The calculated energies are compared with full configuration interaction (FCI) (dot-dash line), and also with CCSD (dotted line), and UCCSD (dashed line) methods. Each of the plots also show the Hartree-Fock (HF) energy as a black solid line, which allows to visualize the correlation effects. 
    } 
\label{fig:s3g}
\end{figure*}

\begin{figure*}[t]
    \centering
    \setlength{\tabcolsep}{1mm}
    \begin{tabular}{ccc}
         \includegraphics[width=60mm]{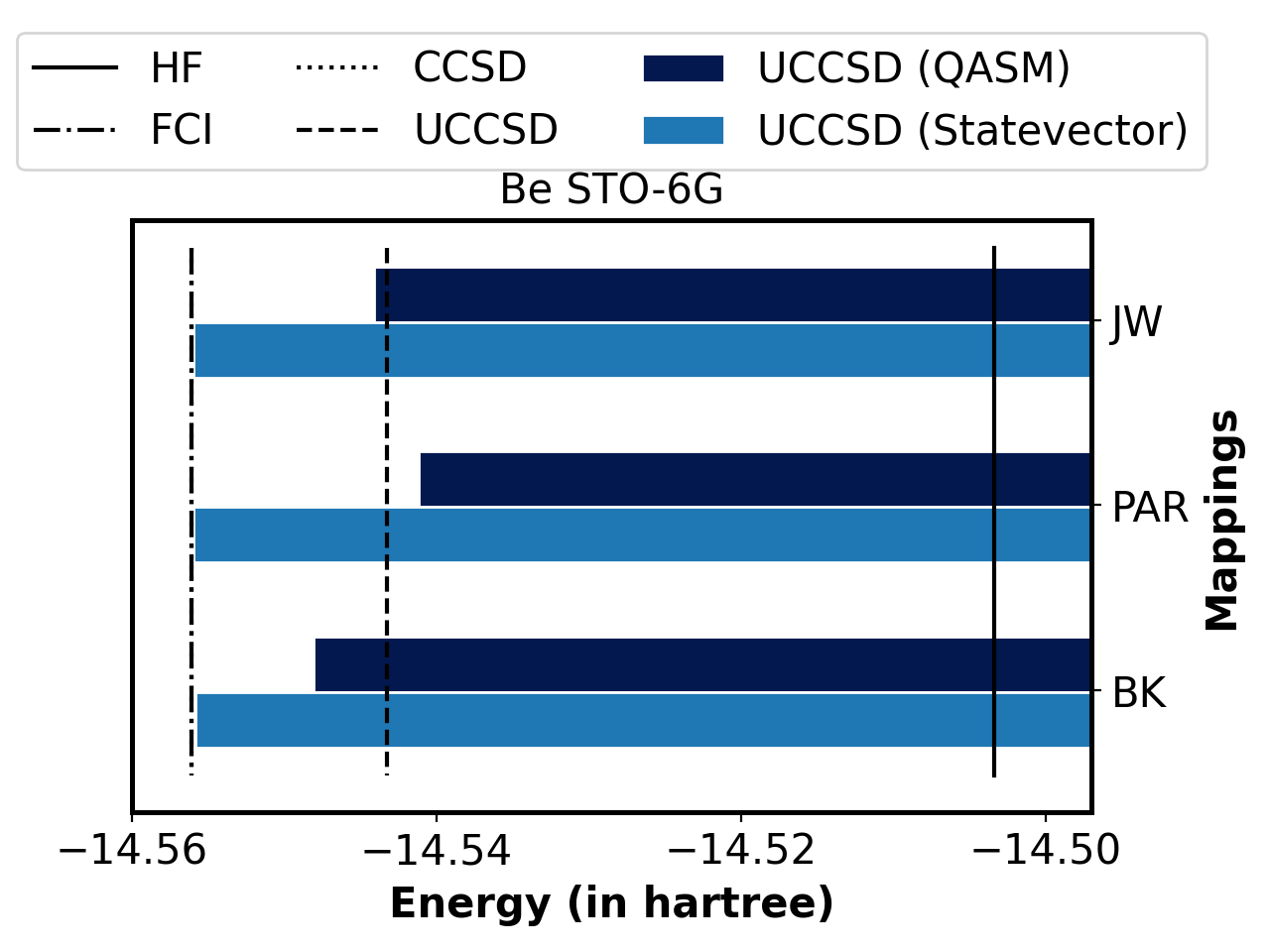}& \includegraphics[width=60mm]{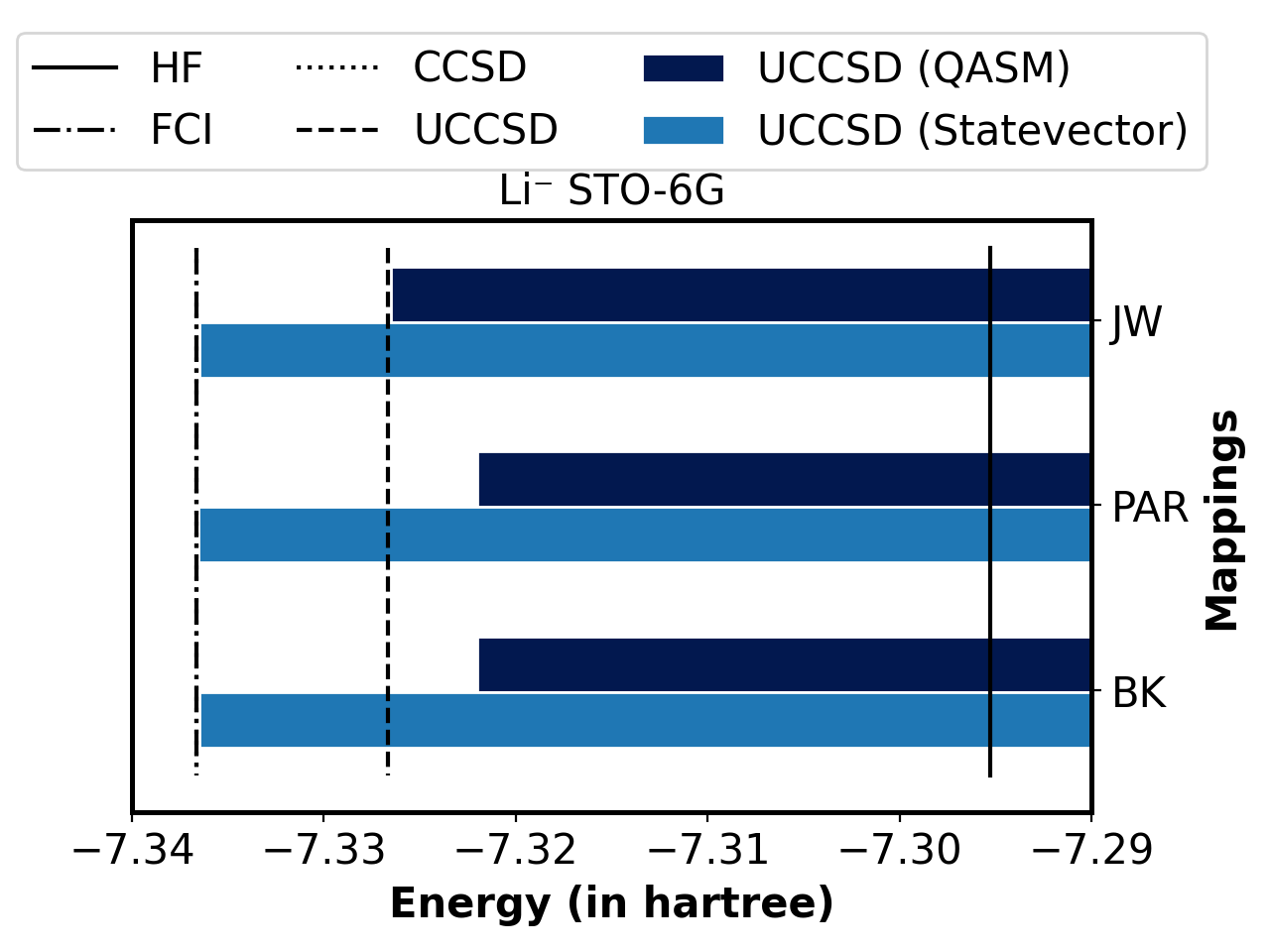} & \includegraphics[width=60mm]{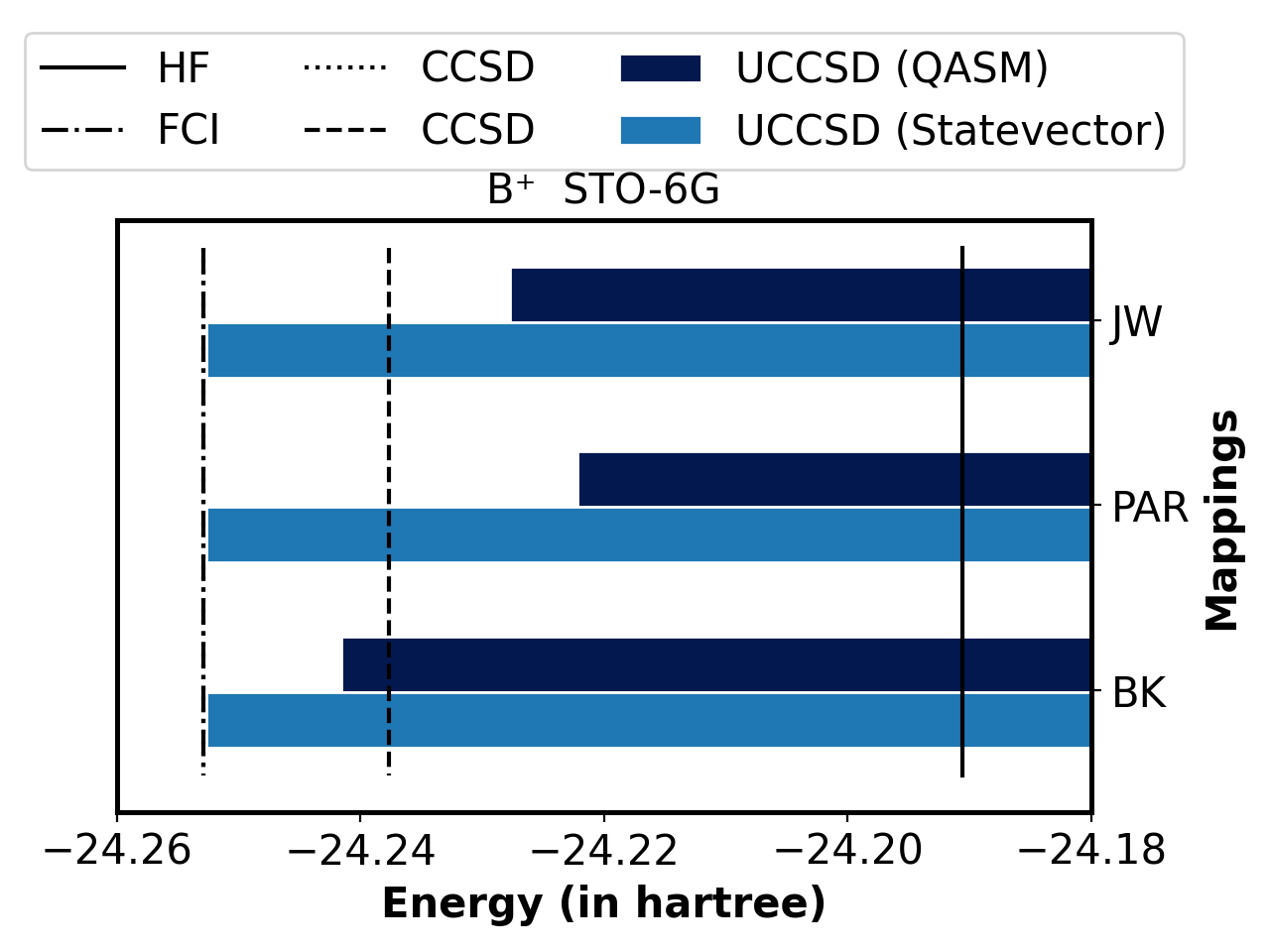}  \\
         \textbf{(a)}& \textbf{(b)} & \textbf{(c)}\\
    \end{tabular}
    
    \caption{Figure showing bar plots depicting the values of ground state energies of the Be, Li$^-$, and B$^+$ obtained from VQE calculations in the STO-6G basis, with various fermion to qubit mapping techniques, and on different backend simulators. The notations are the same as in Fig. \ref{fig:s3g}. } 
\label{fig:s6g}
\end{figure*}

\begin{figure*}[t]
    \centering
    \setlength{\tabcolsep}{1mm}
    \begin{tabular}{ccc}
         \includegraphics[width=60mm]{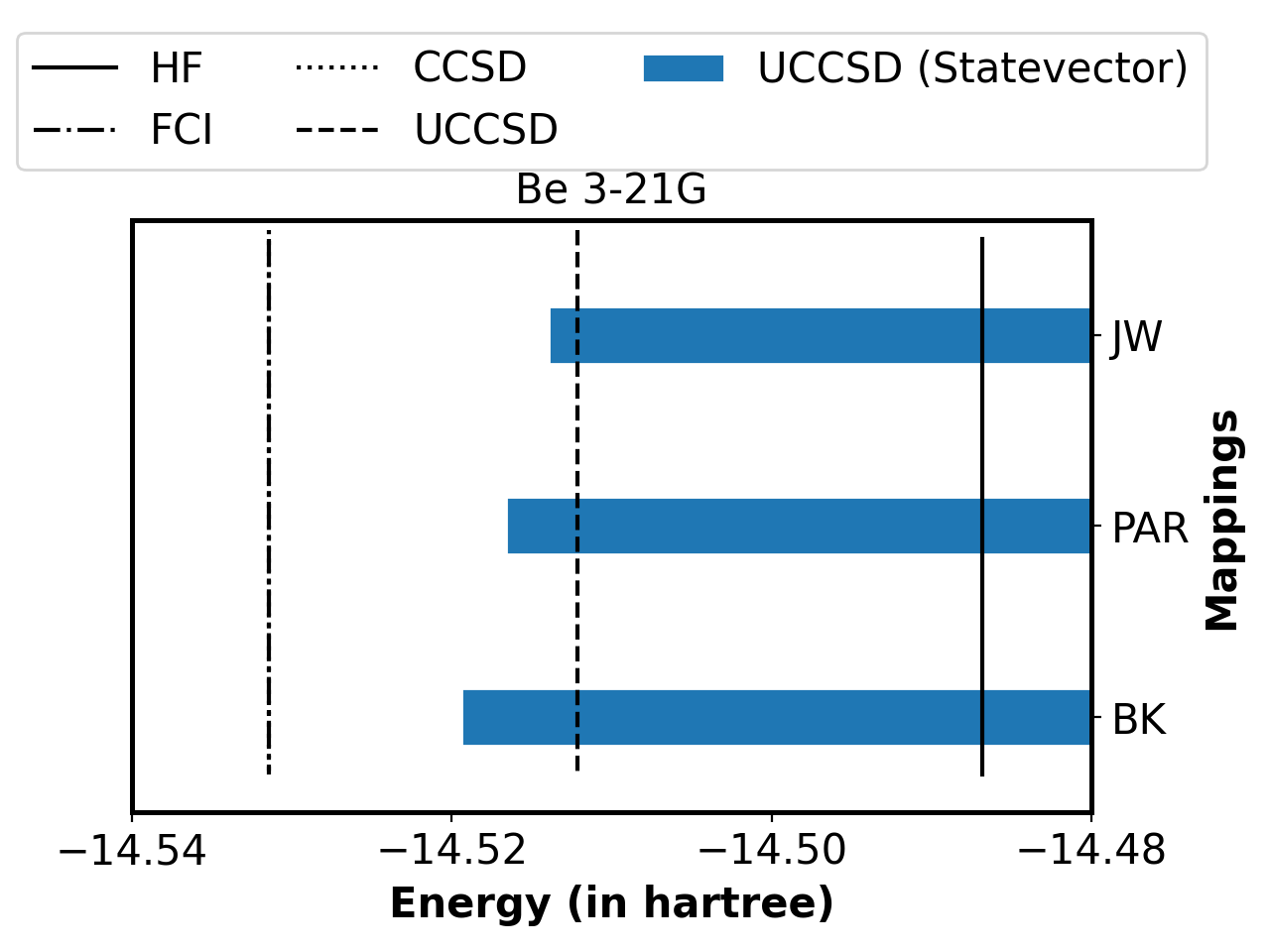}& \includegraphics[width=60mm]{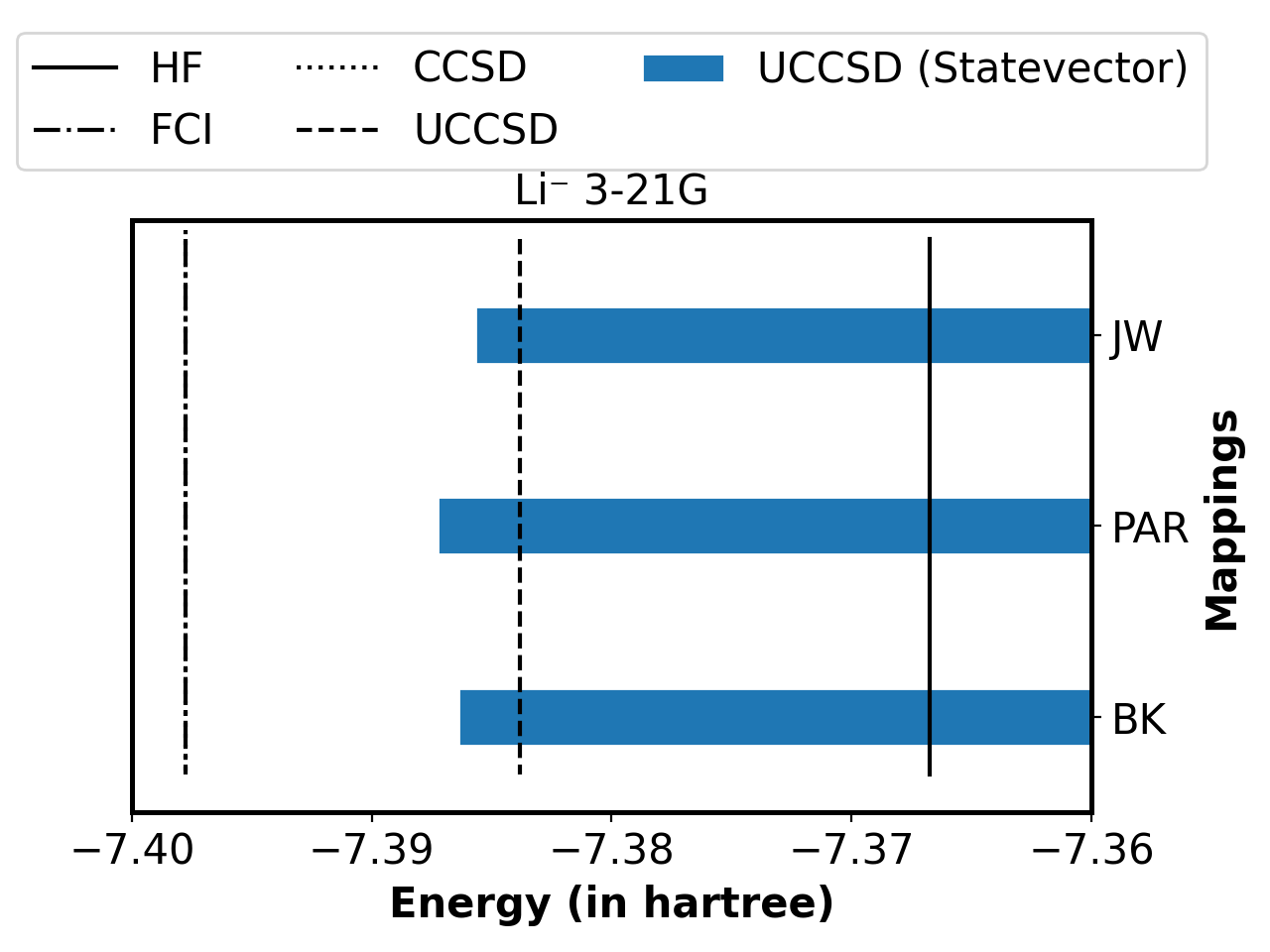} & \includegraphics[width=60mm]{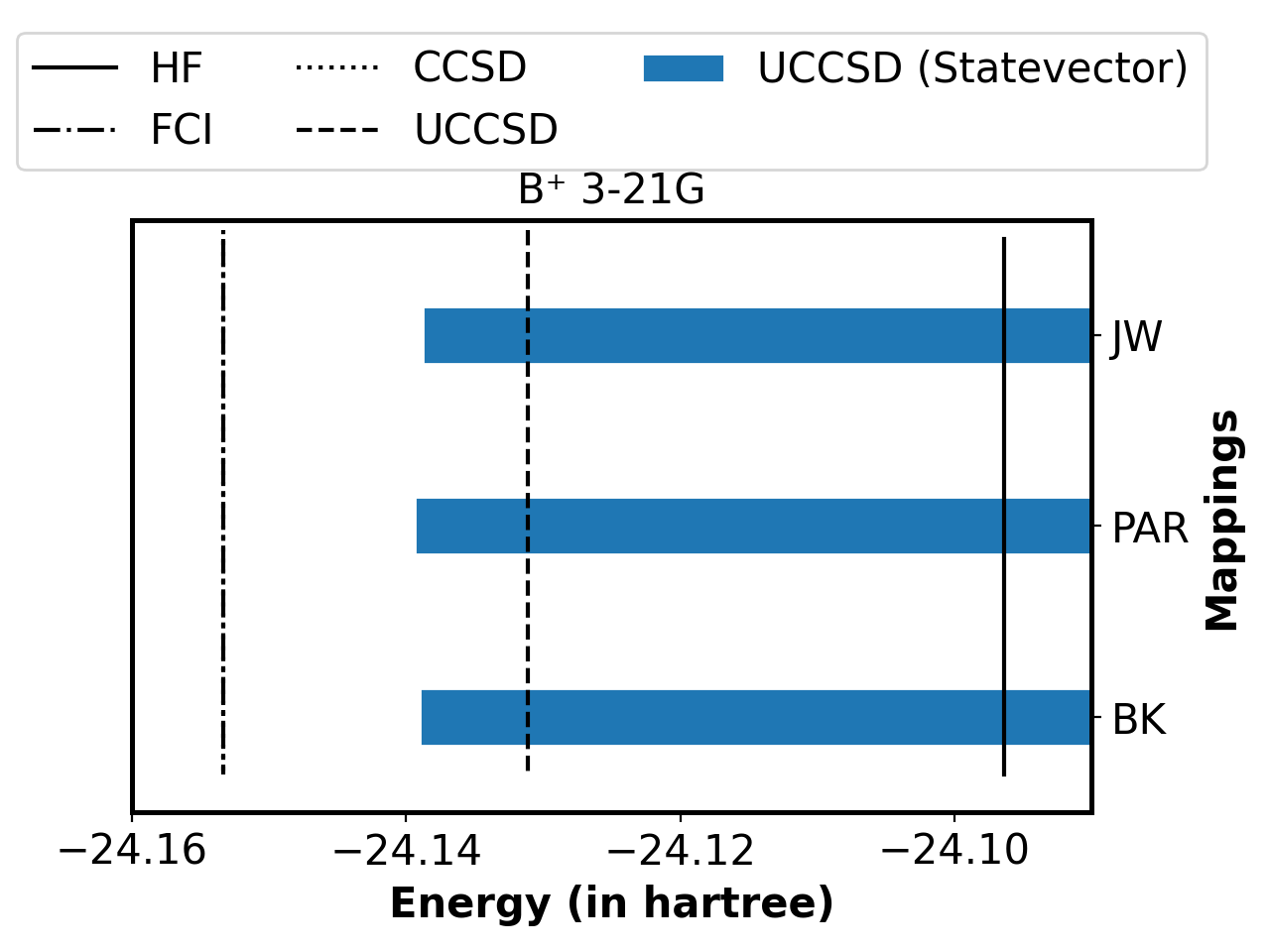}\\
         \textbf{(a)} & \textbf{(b)}&  \textbf{(c)}\\
    \end{tabular}
    \caption{Our results for the ground state energies of the Be, Li$^-$, and B$^+$ in 3-21G basis using different combinations of mappings and backend simulators visualized as bar plots. The notations that have been adopted for this figure are the same as those for Fig. \ref{fig:s3g}.}
 \label{fig:321g}
\end{figure*}

\begin{figure*}[t]
    \centering
    \setlength{\tabcolsep}{1mm}
    \begin{tabular}{ccc}
         \includegraphics[width=60mm]{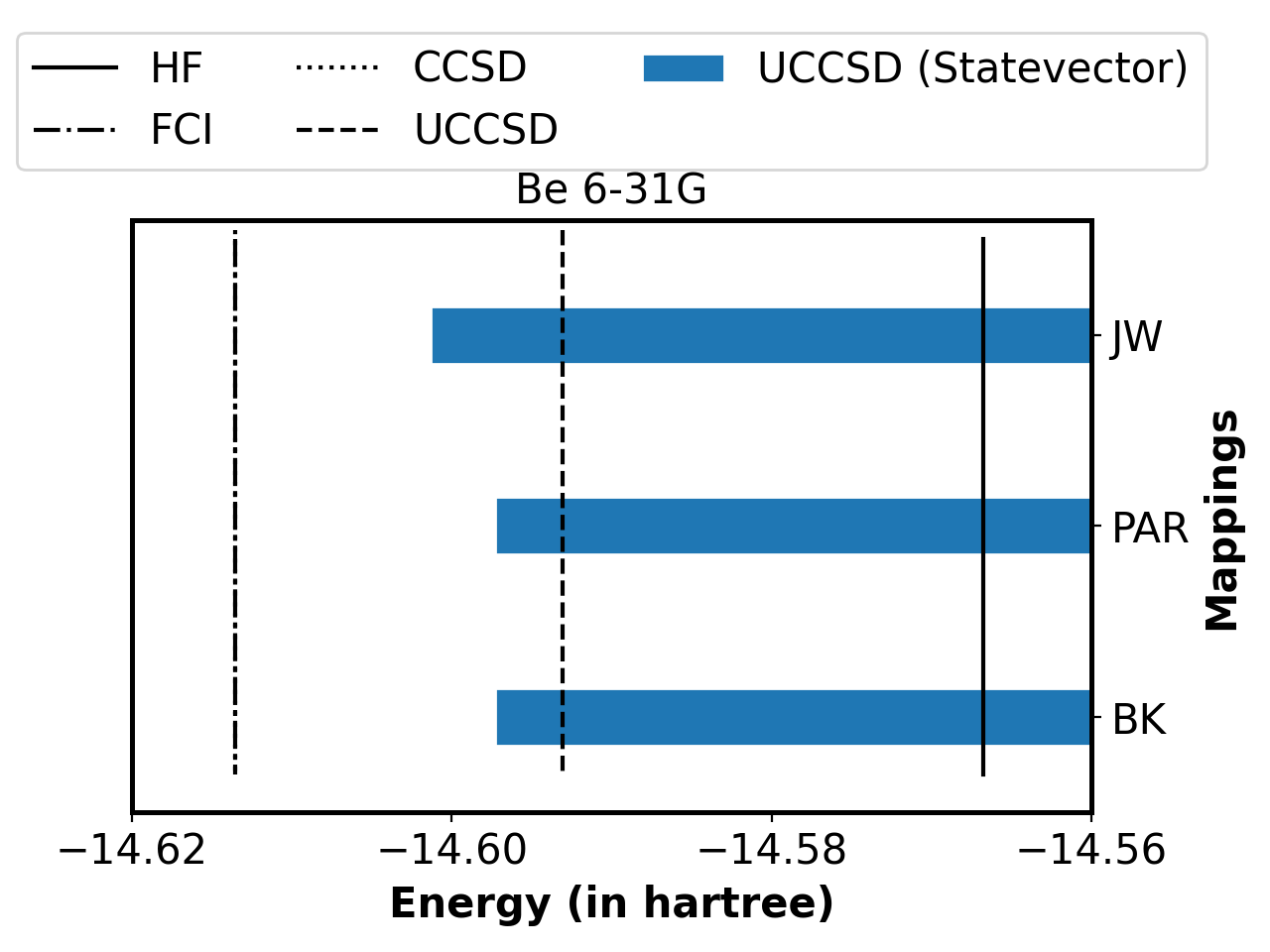}& \includegraphics[width=60mm]{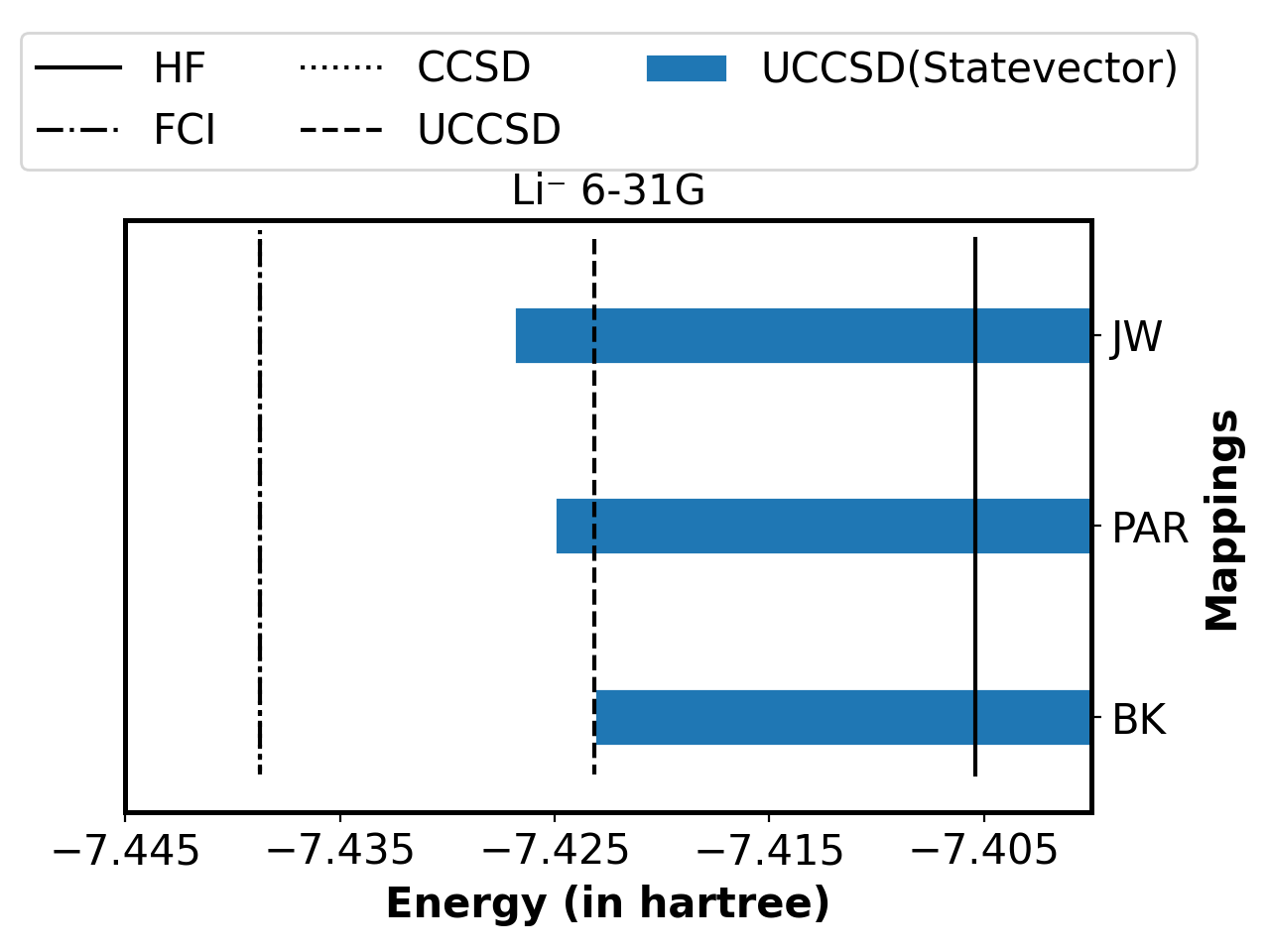} & \includegraphics[width=60mm]{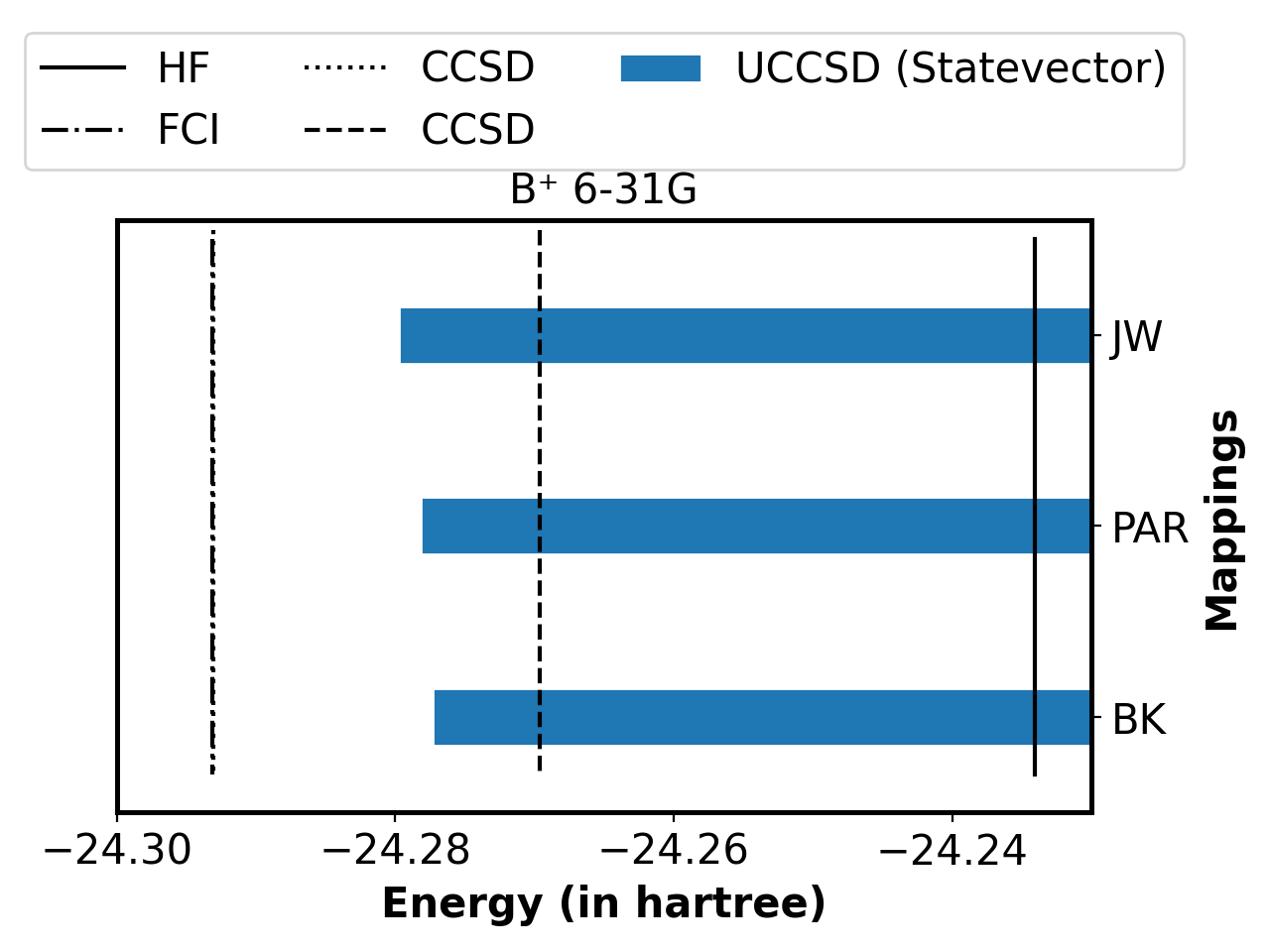}\\
         \textbf{(a)} & \textbf{(b)}&  \textbf{(c)}\\
    \end{tabular}
    \caption{The results for the energies of the Be, Li$^-$, and B$^+$ in the 6-31G basis in the form of bar plots showing different combinations of fermion to qubit mapping techniques and backend simulators. The notations that have been adopted for this figure are the same as those for Fig. \ref{fig:s3g}.}
 \label{fig:631g}
\end{figure*}

\subsection{Analysis of the required number of shots for QASM calculations}

We analysed the number of required shots, for the results obtained using the QASM simulator and chose Be atom as a representative system in the STO-3G basis with JW mapping. The findings from a preliminary analysis of percentage fraction error with respect to FCI versus number of shots, with the latter verified up to 512 shots in steps of one, is given in Fig.~\ref{fig:shots}. We deemed this analysis as being qualitative, in that in a calculation with a given number of shots, the computation does not return identical results when repeated. Hence, we only pay attention to the overall trend for the purposes of this analysis. We note that each point on the X-axis in Fig.~\ref{fig:shots} is an individual computation with those many shots. For 100 shots, we have approximately 4 percent error for Be, which is clearly not desirable. This leads us to Fig.~\ref{fig:25k_shots}, which shows results from larger intervals, and all the way up to 100000 shots. Also, we explicitly plot the energy versus the number of shots here. This analysis is rigorous, with the inclusion of maximum, minimum, and mean values for the energies at each data point, by repeating each of the runs for a given number of shots 160 times. In Fig.~\ref{fig:shots}, we also show the values of energy calculated on a ClC from the HF, FCI, CCSD and UCCSD methods, so as to have a visual feel of correlation effects. It is noticeable from the above figure that at 100 shots, the mean energy is above the HF value, and therefore hardly satisfying the variational principle. One can also see that at lower number of shots, the error bar (the difference between the maximum and minimum values) is so large that its extent is greater than the difference between the energy values from the HF and FCI methods, that is, the amount of electron correlation. As the number of shots increase, the curve approaches and appears to converge to the UCCSD value that one obtains with a ClC and has a very small error bar. It is worth noting here that had we increased the shots further, the curve would have, {\it albeit} gradually, yielded lower values. The inference that the curve would continue to monotonically decrease is based on a simple fit to the mean energy values. However, it is important to see that it is non-trivial to find a rigorous fit due to the statistical nature of each data points, and for our purposes, not necessary. The plot also shows that the error bars reduce with increasing shots. Based on these results, we performed computations with the QASM backend for the rest of the basis sets and mappings, as well as for the other atoms, setting the number of shots to 20000. The rationale is that 20000 shots finds a golden mean between computational cost and accuracy. In fact, we obtained about 0.1 percent error for the mean value with respect to FCI, while the standard deviation is only around 4 mHa. We also add at this point that we carried out bootstrapping procedure to check its agreement of the mean over 160 repetitions for a given number of shots. For the purpose of bootstrapping, we chose 50 samples of lists with each having 4 data points (we did repeat the procedure for longer lists (up to lists of length 50 and from 10 to 100000 shots for each list length), and found that the results only change at $\sim$ 0.1 mHa from 10000 shots onwards). We find that mean and bootstrapped mean agree at 1 mH at 20000 shots. Lastly, in the interest of computational time, we only performed one calculation with 20000 shots for each of the remaining cases (that is, Be with other mappings and also the remaining two atomic systems considered) and not with twenty repetitions, given that for Be with STO-3G basis and JW mapping, at those many shots, the difference between the maximum and minimum values in twenty repetitions is less than 0.1 percent. We anticipated the error estimate to be similar for the rest of the cases, and under this assumption, we performed a single run for them. It is worth adding at this point that there are works in literature that have made great strides in reducing the number of VQE measurements for electronic structure~\cite{shots1,shots2,shots3,shots4}. Lastly, we note that this analysis serves an important purpose; our estimate for the number of shots required sets the tone for future analyses on atomic systems with the QASM backend, where we can strive to emulate a quantum computer more realistically, with the inclusion of noise models and error mitigation. \\

\subsection{Analysis of errors due to Trotter number}

We verified the errors that may arise as result of setting the Trotter number in our simulation to one. For the Be atom in the STO-3G basis and with JW mapping, we found that up to a Trotter step of 50, the error was at most $\sim$1 milli-hartree in $\sim$14 hartree. For Li$^-$ with the same basis and mapping, the error was found to be as high as 0.2 milli-hartree in $\sim$7 hartree, and for B$^+$, the error did not exceed 0.1 milli-hartree in $\sim$24 hartree. We also verify that we obtain similar estimates with Parity and BK mappings for all three chosen atomic systems. Hence, we set a conservative estimate that even with other basis sets and mappings, the error due to Trotter step would not exceed 0.01 percent. In other words, the error is negligible, and justifies setting Trotter number to one. We note that trottering $U(\theta)$ does not preserve particle number, and in that sense, one can view the comparison between VQE results and FCI results as the deviation in our results from the expectation value of the correct particle number. 

\subsection{Main results and analysis}

We examine the correlation effects in the ground state energies of the systems that we considered, in Tables~\ref{tab:s3g} through \ref{tab:631g}, with each table presenting results for a given basis set. Among these, Table \ref{tab:s3g} (and the accompanying Fig. \ref{fig:s3g}) gives the STO-3G results. We immediately see that for Be, the energies obtained using the statevector simulator agree to $\sim$ 0.1 milli-hartree, or about 0.001 percent error, with respect to FCI. We find similar differences for Li$^-$ and B$^+$ for the STO-3G basis, whose results are also presented in Fig. \ref{fig:s3g}. In comparison, the correlation effects from FCI are about 50, 40, and 60 milli-hartree for Be, Li$^-$, and B$^+$, respectively. Therefore, we can infer that quantum simulation with statevector simulator accounts for electron correlations very accurately in the STO-3G basis. This is perhaps not surprising, as a statevector simulator does not rely upon statistics built from repeated measurements in order to extract energy. We also present our results from a QASM simulator. They are all in good agreement with the UCC results from a ClC, and not with the FCI results as one may expect, due to our choice of the number of shots (20000 of them) as seen earlier. In fact, the difference between the FCI and the UCC QASM value (about 8 and 11  milli-hartree for PAR and BK mappings, respectively, and about 15 milli-hartree for JW case), is comparable with that between the maximum and the minimum values (about 8 milli-hartree) for 20000 shots from Fig.~\ref{fig:25k_shots}. \\

A peculiar observation in the ClC part of the results is that for all the considered basis sets, the CCSD result agrees better with the results from the FCI method than the UCCSD method. In principle, UCCSD is expected to capture more many-body effects than CCSD, with the caveat that the energy expression for the former does not naturally terminate, thereby relying upon the chosen truncation scheme to achieve the desired results. We suspect that the observed deviation is associated with the truncation scheme of the UCCSD method. To that end, we provide a fairly detailed explanation of the approach taken in the openfermion code to evaluating the UCCSD energy, and en route, explain the approximations and truncation involved in the procedure. The UCCSD energy is calculated using the familiar $E = \langle \Phi_0 | e^{\Theta \dag}H e^\Theta | \Phi_0 \rangle$. Rather than taking the conventional route of solving the UCCSD equations, obtain the amplitudes, and then solve the expression for energy while suitably terminating the series, the code aims at directly solving the expression for energy by matrix multiplications involving $H$ and $e^\Theta | \Phi_0 \rangle$. In order to do this, the second quantized creation and annihilation operators in the Hamiltonian and wave function are mapped to a string of tensor products of Pauli operators using the Jordan-Wigner transformation. The CCSD equations are solved in Pyscf or any other suitable program, and the t amplitudes (real) thus obtained are used in $e^{\Theta}$. The remaining step, and arguably the most important, is finding an efficient way of matrix multiplication. This boils down further to finding an efficient approach for multiplying an exponential matrix and a column vector. The algorithm from Al-Mohy and Higham~\cite{almohy} is adopted, where $e^{tA}B$, where $A$ is an $n \times n$ matrix, and $B$ an $n \times n_0$ matrix, with $n_0 << n$, is evaluated by \textit{approximating} the exponential as a $[m/m]$ Pade approximant, which in turn is written as a \textit{truncated} Taylor series. The backward error from such an algorithm would depend on the choice of an integer, $s \geq 1$, and $m$, which are in turn chosen by an appropriate recipe from their paper. On the other hand, the UCCSD energy obtained from qiskit simulators (where we have a compact circuit representation for exponentials, after Trotterization), $E = \langle \Phi_0 | e^{-\Theta}H e^\Theta | \Phi_0 \rangle$, have their own sources of errors, which we have studied in detail in our work. Given that the approaches are different, complex, and come with their own approximations, and also given that evaluating a matrix exponential is almost always non-trivial, a thorough study of the openfermion UCCSD algorithm over and above the details that we mentioned above is perhaps beyond the scope of this work. To that end, we have done the best that we can, which is to compare both the UCCSD results with the FCI result with the same single-particle basis. \\

Table \ref{tab:s6g} (and Fig. \ref{fig:s6g}) also shows the same results but obtained with the STO-6G basis. The results are an improvement over the earlier basis as evident by lowering of the calculated energies, although the qubit number is the same for a given system, since more functions are contracted in the STO-6G case. Not too surprisingly, the trends are very similar to those in the STO-3G basis. \\


We now proceed to examine the results obtained from  bigger bases as shown in Tables \ref{tab:321g} and \ref{tab:631g} (and the accompanying Figs. \ref{fig:321g} and \ref{fig:631g} respectively). We reiterate that QASM results are not computed for calculations using these two basis sets, in view of the requirement of a large number of shots to obtain a reasonably accurate result. We observe from the Table \ref{tab:321g} that the effect of electron correlation on FCI energy is about 40, 30, and 50 in milli-hartree for Be, Li$^-$, and B$^+$, respectively, whereas the difference in the correlation energies between FCI and quantum simulation are about 10 milli-hartree for all the systems. This discrepancy is possibly due to the slow convergence of the COBYLA optimizer. To check this, we choose the JW mapping and the STO-3G basis set for a representative calculation, and increase the number of iterations to beyond the default maximum threshold of 1000 iterations (which we employ to report our results in this work). We found that while the percentage fraction error with respect to the FCI result is $\sim10^{-3}$ at 1000 iterations, it decreases further to $\sim10^{-4}$ at 2000 iterations. We expect that with the 3-21G basis as well as the 6-31G basis, the results would improve slightly with larger number of iterations, which comes with higher computational cost. Alternatively, one could employ an optimizer that converges faster, such as L-BFGS-B and conjugate gradient, which we find after a preliminary survey to have converged within a lesser number of iterations but not as smoothly as COBYLA. \\

\subsection{Further findings from obtained data}

In the rest of the results section, we briefly present an assortment of important comments based on our findings.\\

\textit{Dependence of results on mapping scheme: }It is known that the energy landscape close to the variational minimum is invariant with respect to the chosen encoding. It is, therefore, important to note that the observed difference in results across maps for the QASM simulator could be more a consequence of the error due to statistics associated with the backend, and not the actual difference, if any. En route to drawing this conclusion, we have explicitly verified that the errors due to trotterization and optimizer convergence have been checked and are found to be negligible, by choosing Be as a representative case and for all the three mapping schemes. \\

\textit{Preferred mapping scheme: }We note that for a given atom, between different maps, the change in correlation energies even with the QASM backend is $\sim$1 milli-hartree, thus reinforcing that the correlation energy is not very sensitive to the mapping scheme. In this regard, the PAR map is cheaper due to the reduction of two qubits, while giving results in agreement with other maps that are more qubit-expensive, and hence recommended for future atomic calculations of this nature. \\

\textit{STO-6G vs 3-21G bases: }The largest basis chosen in this work, namely the 6-31G basis, displays trends similar to the 3-21G counterpart. An observation about the results from the 3-21G basis is that the obtained FCI results (and hence, statevector results and predicted QASM results with 20000 shots) are comparable to those from the STO-6G basis for Be (within 10 milli-hartree), whereas the 3-21G results are slightly better (about 60 milli-hartree) and much worse (about 100 milli-hartree) for the negative and positive ions, respectively, than the STO-6G basis. However, since the STO-6G basis uses 10 qubits while 3-21G demands 18 for the considered systems, the former is more attractive and should be preferred over the latter.\\ 

\textit{Trends in C$^{2+}$: }Since our goal in this pilot study is to assess precision with which one can capture correlation effects, and not really present an exhaustive survey of many systems, we have tried to select carefully a few systems and attempted to be rigorous in studying the influenze of the knobs of the VQE algorithm in deciding the precision in ground state energies. As an extension, we also comment on two other isoelectronic systems to Be. We explicitly verify that the trends, as one would expect, remain the same in C$^{2+}$ by calculating its energies within the STO-3G and STO-6G bases, for all three mappings, and compare with FCI, CCSD, and the UCCSD results from a classical algorithm, just as we did for the main systems considered in this work. Within each of the two bases, our statevector (SV) results differ at $\sim$ 0.1 mHa with respect to the FCI values, while for QASM backend (with 20000 shots), they differ at $\sim$ 10s of mHa. We add that we have avoided the isoelectronic He$^{2-}$, as the STO series contracted basis sets have been designed in a way historically where there is no room for any correlation effects.\\

\textit{VQE results versus FCI in larger bases: }Lastly, we attempt to address the question of the energy from a VQE calculation being farther away from the FCI value in larger basis sets, as compared to the smaller ones. This could be due to the fact that with lesser qubits in a smaller basis and hence a limited number of virtuals, we miss a fewer excitations, whereas for a larger basis with more virtuals, we miss more excitations from the higher-level excitations such as from the triples and quadruples. \\

\section{Conclusion}

We investigated the trends in electron correlation effects for assessing the precision with which we can determine the ground state energies of Be, Li$^-$, and B$^+$ isoelectronic systems using the quantum-classical hybrid variational quantum eigensolver algorithm. We worked with four single-particle basis sets, two minimal and two split-valence, that are not very qubit-expensive and thus suited for the Noise Intermediate State Quantum era. Within each of those bases, we analyzed the changes in the results with choice of mapping as well as the backend simulator, in the unitary coupled-cluster theory ans\"{a}tz. The energies obtained using the STO-3G basis showed that the statevector results agreed to well beyond the milli-hartree level of precision with respect to the best possible calculation within that single-particle basis, namely the full configuration interaction method, for all the three atomic systems. Moreover, the results were found to be almost independent of the choice of mapping. For calculations using the QASM simulator backend, we first carried out an extensive analysis of the required number of shots, and with our recommended value, the results agreed to tens of milli-hartree with respect to the full configuration interaction method. We also probed the errors due to the spread in results from repetition of a computation with a given number of shots, and found that they lie around the same ballpark, and the energies are in better agreement with the values obtained using the singles and doubles approximated unitary coupled-cluster theory on a classical computer. The results from the higher quality STO-6G basis also showed similar trends. When we examined the statevector results from the larger (in number of qubits) 3-21G basis set, we found that the values were comparable with those obtained using the STO-6G basis set and with the same backend, although the former required eight extra qubits than the latter. We find again that the results from the 3-21G basis agree better with the singles and doubles approximated unitary coupled-cluster theory. Also, the results vary within 10 milli-hartree across different mappings. The trends from the 6-31G basis are  similar to those from the 3-21G basis set. Our work is timely and relevant in view of recent developments, such as the announcement of the IBM Quantum Condor, which is expected in 2023, which aims to hit the 1000-qubit mark in the quest for scaling quantum computers~\cite{gambetta}. When such machines come to fruition, our pilot study will pave the way in  extending our present calculations to not only heavier atomic systems, but also in evaluating other properties with the VQE algorithm, which would be of interest in new physics beyond the Standard Model and other important applications such as atomic clocks. 

\section*{Acknowledgements}
We thank Mr. Ramanuj Mitra and Dr. Amar Vutha for their help with computational resources for the calculations reported in this work. We are also grateful to Dr. Kenji Sugisaki for useful discussions. Sumeet thanks Physical Research Laboratory (PRL), Ahmedabad for providing the visiting fellowship to carry out a part of this work. Most of the computations were performed on the VIKRAM-100 cluster at PRL and CDAC's PARAM Siddhi machine. We acknowledge National Supercomputing Mission (NSM) for providing computing resources of `PARAM Siddhi-AI' at C-DAC Pune, which is 
implemented by C-DAC and supported by the Ministry of Electronics and Information Technology (MeitY) and Department of Science and Technology (DST), Government of India.  V.S.P. acknowledges the Graham cluster at the SciNet HPC Consortium (Compute Canada). SciNet is funded by: the Canada Foundation for Innovation; the Government of Ontario; Ontario Research Fund - Research Excellence; and the University of Toronto. V.S.P. and Sumeet both used Google Colab (Bisong E. (2019) Google Colaboratory. In: Building Machine Learning and Deep Learning Models on Google Cloud Platform. Apress, Berkeley, CA. $\mathrm{https://doi.org/10.1007/978-1-4842-4470-8\_7}$) for less compute intensive analyses.

\end{document}